\documentclass[journal]{IEEEtran}
\usepackage{booktabs} % For formal tables

\usepackage[font={small}]{caption}

\usepackage{xcolor,colortbl}  
\definecolor{D}{RGB}{0,0,0} 
\definecolor{CB}{RGB}{18,20,86} 
\definecolor{CS}{RGB}{174,54,0} 
\definecolor{V}{RGB}{3,72,23} 
\definecolor{P}{RGB}{191,5,7} 
\definecolor{darkyellow}{RGB}{225, 170, 7}

\usepackage{pgfplots}  
\usepackage{tikz}
\usetikzlibrary{shapes.geometric}
\usetikzlibrary{shapes.misc, positioning}
\usepackage{amsmath}
\usepackage{url}
\usepackage{enumitem}
\setlist[enumerate]{itemsep=0mm}

\usepackage{multirow}

\newcommand*{\ie}{{\em i.e.,~}}
\newcommand*{\eg}{{\em e.g.,~}}
\newcommand*{\etal}{{et al.~}}
\newcommand*{\ignore}[1]{} 
\newcommand*{\reduceSpaceBeforeCaption}{\vspace{-.2in}} 
\newcommand*{\reduceSpaceAroundFigure}{\vspace{-.1in}} 

\newcommand*{\rqone}{RQ1: How evenly do users share content from news sources of varying credibility?} 
\newcommand*{\rqtwo}{RQ2: How many users share content from different types of news sources?}   
\newcommand*{\rqthree}{RQ3: How quickly do users share from news sources of each type?} 
\newcommand*{\rqfour}{RQ4: Who is sharing from different news sources?}  

\newcommand*{\rqonesection}{\medskip \noindent \textbf{\rqone}  \smallskip \\  \noindent }

\newcommand*{\rqtwosection}{ \medskip \noindent \textbf{\rqtwo}   \smallskip \\  \noindent }

\newcommand*{\rqthreesection}{ \medskip \noindent \textbf{\rqthree}   \smallskip \\ \noindent }

\newcommand*{\rqfoursection}{ \medskip \noindent \textbf{\rqfour}  \smallskip \\ \noindent }

\usepackage{amsthm}

\begin{document}
\title{Propagation from Deceptive News Sources\\{\huge Who Shares, How Much, How Evenly, and How Quickly?}} 

\author{Maria~Glenski, %~\IEEEmembership{~}
        Tim~Weninger, %~\IEEEmembership{~}
        and~Svitlana~Volkova%,~\IEEEmembership{~}% <-this % stops a space
\thanks{Manuscript received March 23, 2018, revised July 20, 2018, accepted September 24, 2018. The research described in this paper was supported by the
Laboratory Directed Research and Development Program at Pacific Northwest National Laboratory, a multiprogram national laboratory operated by Battelle for the U.S. Department of Energy under contract \#286504 and DARPA under USC subcontract \#94725237.} 
\thanks{\textit{Corresponding authors: T. Weninger and S.Volkova.}}
\thanks{M. Glenski and T. Weninger are with the Department
of Computer Science and Engineering, University of Notre Dame, Notre Dame, IN 46556 USA e-mail: (\{mglenski, tweninge\}@nd.edu).}% <-this % stops a space
\thanks{S. Volkova is with Data Sciences and Analytics Group, National Security Directorate,
Pacific Northwest National Laboratory, Richland, WA 99354 USA email: (svitlana.volkova@pnnl.gov).}}% <-this % stops a space

\markboth{~}%IEEE Transactions on Computational Social Sciences,~Vol.~1, No.~1, January~2999}%
{Glenski \MakeLowercase{\textit{et al.}}: News Propagation and Influence from Deceptive Sources}


\maketitle

\begin{abstract}
		As people rely on social media as their primary sources of news, the spread of misinformation has become a significant concern. In this large-scale study of news in social media we analyze eleven million  posts and investigate propagation behavior of users that directly interact with news accounts identified as spreading trusted versus malicious content. Unlike previous work, which looks at specific rumors, topics, or events, we consider all content propagated by various news sources. Moreover, we analyze and contrast population versus  sub-population behaviour (by demographics) when spreading misinformation, and distinguish between two types of propagation, \ie  direct retweets and mentions. Our evaluation examines how evenly, how many, how quickly, and which users propagate content from various types of news sources on Twitter. 
		
		Our analysis has identified several key differences in propagation behavior from trusted versus suspicious news sources. These include high inequity in the diffusion rate based on the source of disinformation, with a small group of highly active users responsible for the majority of disinformation spread overall and within each demographic. 
		Analysis by demographics showed that users with lower annual income and education share more from disinformation sources compared to their counterparts. News content is shared significantly more quickly from trusted, conspiracy, and disinformation sources compared to clickbait and propaganda. Older users propagate news from trusted sources more quickly than younger users, but they share from suspicious sources after longer delays. Finally, users who interact with clickbait and conspiracy sources are likely to share from propaganda accounts, but not the other way around.  
\end{abstract}

\begin{IEEEkeywords}
social network analysis, information propagation, disinformation, misinformation, deception.
\end{IEEEkeywords}

	\section{Introduction}
	\noindent People use social media not only for entertainment and social networking but also as their primary source of news and information. An August 2017 survey from the Pew Research Center found that 67\% of Americans report that they get at least some of their news from social media, an increase of 5\% over the previous year~\cite{shearer2017pew}. Of those who use Twitter, 74\% said they received news from the platform. With such a reliance on social media as a source of news and information, the spread of misinformation is a significant concern.

	Previous work in social media analysis, especially within the area of social news, has focused on influence campaigns and the spread of (mis)information either organically or through bots. Given a particular event like an election or a natural disaster, researchers typically follow information cascades to tease out diffusion processes and infer various characteristics about how social media responded to the event~\cite{takahashi2015communicating,ferrara2017disinformation}. These studies have resulted in important findings about the effect of such items as information contagion~\cite{lerman2010information}, influence campaigns~\cite{glenski2017rating}, bots~\cite{ferrara2016rise}, and spam~\cite{hadgu2013political}, etc., within specific newsworthy events.

	In the present work we take a different view. Rather than studying information propagation one newsworthy event at a time, we seek to quantify and compare socio-digital phenomena \textit{according to the source of the information}. 
	This is especially prescient in the current climate where the reliability of traditional sources of news and information are contested. For this we rely on previous work by Volkova et al. \cite{volkova2017separating} that aimed to classify information sources according to their quality (\ie their accuracy according to fact-checking organizations) and their intent (\ie whether the author intends to deceive the reader or not). Along these two axes we focus on news sources that fall into one of the following five categories:
	trusted, clickbait, conspiracy theories, propaganda, and disinformation.

	The goal of the present work is to quantify and compare the immediate propagation of information from the different types of news sources. In service of this goal we identified  282 news sources on Twitter and collected 11 million direct interactions (\ie retweets and mentions) with those source accounts from two million unique users. Unlike previous work, we report our findings on {information propagation behaviour from sources of varying levels of credibility at the population level as well as for various user-demographics}. With this data and the news source type classification, which is described in more detail in the next section, we can quantify how social media users interact with different types of news sources. This is the focus of the four research questions outlined below.

	\smallskip\noindent\textbf{\rqone} 
	
	Several previous studies have investigated the makeup of users that retweet content from specific news sources as a way to identify sources that spread rumors or disinformation. In the context of social media, the 1\% rule and its variants indicate that most users only browse content while a mere 1\% of users contribute new content~\cite{vanmierlo2014rule,hargittai2008participation}. Within the subset of those who actively contribute new content, Kumar and Geethakumari \cite{kumar2014detecting} found larger disparity among users who retweeted news from sources identified as spreading disinformation. That is, a small number of highly active users were responsible for the vast majority of retweets of disinformation. However, fitting the template of most social network research, the study focused only on keywords related to the events in Egypt and Syria in 2013. To answer this research question more generally, the present work quantifies and compares the disparity in sharing behavior of millions of users across the various categories of news sources. Specifically, for each type of news source --- clickbait, propaganda, etc., we ask whether information sharing is equally distributed across users, or instead if there are a small group of vocal users responsible for the majority of the information propagation.

	\medskip\noindent\textbf{\rqtwo}
	
	To identify rumor-spreading users, Rath et al. \cite{rath2017retweet} proposed an RNN model with believability scores to weight edges in a user-retweet network. Believability scores for pairs of users were calculated from the users' scores of trustingness and trustworthiness. They used the propensity of other users to retweet a source as a proxy for the trustworthiness of the source. Users were considered to be more trustworthy if more users retweeted them, and users were considered more trusting if they retweeted content from a larger number of other users. Using a similar proxy for trustworthiness, we consider whether deceptive sources can be identified by how trustworthy they are, \ie how many users retweet their posts.

	\medskip\noindent\textbf{\rqthree}
	
	Information diffusion studies have often used epidemiological models to understand the diffusion of information, both suspicious and trusted, among social media users \cite{jin2013epidemiological,tambuscio2015fact,wen2015sword,wu2016mining}. These models were originally formulated to model the spread of disease within a population. In the social media context, users are considered to be ``infected'' when they propagate information to other users. 
	Jin et al. \cite{jin2013epidemiological} modeled diffusion using the SEIZ (Susceptible, Exposed, Infected, and Skeptics) model and compared ratios of the transition rates into and out-of their ``exposed'' category, \ie whether people are exposed to misinformation faster than they spread it.  
	Authors found that users tend to share information about factual events more quickly than misinformation or rumors (both verified false or ambiguous in veracity). A recent study by Vosoughi \etal~\cite{Vosoughi1146} found that news fact-checked and found to be false spread faster and to more people than news items found to be true.  Because our methodology considers all content directly shared from various sources (rather than content about specific events), we are able to determine whether deceptive or trusted \textit{sources} have slower immediate share-times overall. 
		
	\begin{figure*}[t!]
		\centering
		\begin{tikzpicture}
\begin{axis}[ 
height=2.8in, width=7in,
legend style={at={(0.75,.25)},
	anchor=north,legend columns=1,column sep=.1cm,draw=none},
xlabel={\small \% tweets of type (log)},
ylabel={\small \# tweets (log)},
xlabel style= {yshift = .1in},
ylabel style= {yshift = -.2in},
font = \small,
xmin=0.0002,%15,
xmax=1.4,ymin=5,ymax=5000000, 
xmode=log,ymode=log, 
xtick = {0.0001,0.001,0.01,0.1,1}, xticklabels = {0.01,0.1,1,10,100},
]

% VERIFIED 

\addplot[mark=o,V,only marks] coordinates {
	(0.1171,802115) %% el_pais
	(0.1149,787140) %% nytimes
	(0.0706,483656) %% Milenio
	(0.0698,477966) %% timesofindia
	(0.0473,324243) %% washingtonpost
	(0.044,301324) %% htTweets
	(0.0381,261050) %% TIME
	(0.0369,252641) %% guardian
	(0.0349,239204) %% Independent
	(0.0345,236238) %% ZeeNews
};
\draw[V] (axis cs:(0.1171,802115)  -- (axis cs:(0.1,3000243) node[anchor=east,rotate=0] { {el\_pais} };
\draw[V] (axis cs:(0.1149,787140) -- (axis cs:(0.1,1500324) node[anchor=east,rotate=0] { {nytimes} };
\draw[V] (axis cs:(0.07,483656) -- (axis cs:(0.035,3000243) node[anchor=east,rotate=0] { {Milenio }};
\draw[V] (axis cs:(0.0698,477966)  -- (axis cs:(0.04,1500324) node[anchor=east,rotate=0] { {timesofindia} };
\draw[V] (axis cs:(0.0473,324243) -- (axis cs:(0.0061,2500243)node[anchor=east,rotate=0] { {washingtonpost} };
\draw[V] (axis cs:(0.044,301324) -- (axis cs:(0.0061,1200324) node[anchor=east,rotate=0] { {htTweets }};
\draw[V] (axis cs:(0.0381,261050)  -- (axis cs:(0.001,1200324) node[anchor=east,rotate=0] {{ TIME} };
\draw[V] (axis cs:(0.0369,252641) --  (axis cs:(0.001,500324) node[anchor=east,rotate=0] { {guardian} };
\draw[V] (axis cs:(0.0349,239204) -- (axis cs:(0.001,240000)  node[anchor=east,rotate=0] { {Independent} };
\draw[V] (axis cs:(0.0345,236238) -- (axis cs:(0.001,110238) node[anchor=east,rotate=0] { {ZeeNews} };

% CLICKBAIT 

\addplot[mark=square,CB,only marks] coordinates {
	(0.4304,17369.0) %% theblaze
	(0.3366,13584.0) %% politicususa
	(0.0942,3802.0) %% veteranstoday
	(0.0744,3004.0) %% theinquisitr
	(0.0311,1255.0) %% yournewswire
	(0.029,1169.0) %% lewrockwell
	(0.0024,97.0) %% other98
	(0.0008,34.0) %% pakalert
	(0.0006,26.0) %% coasttocoastam
	(0.0003,12.0) %% anonews_co
};
\draw[CB] (axis cs:(0.4304,17369.0) node[anchor=west,rotate=0] { theblaze };
\draw[CB] (axis cs:(0.3366,13584.0)  -- (axis cs:(0.4,8000.0)node[anchor=west,rotate=0] { politicususa };
\draw[CB] (axis cs:(0.0942,3802.0) node[anchor=west,rotate=0] { veteranstoday };
\draw[CB] (axis cs:(0.0744,3004.0) -- (axis cs:(0.12,1500.0)  node[anchor=west,rotate=0] { theinquisitr };
\draw[CB] (axis cs:(0.0311,1255.0) node[anchor=west,rotate=0] { yournewswire };
\draw[CB] (axis cs:(0.029,1169.0) node[anchor=east,rotate=0] { lewrockwell };
\draw[CB] (axis cs:(0.0024,97.0) node[anchor=west,rotate=0] { other98 };
\draw[CB] (axis cs:(0.0008,34.0) node[anchor=west,rotate=0] { pakalert };
\draw[CB] (axis cs:(0.0006,26.0) -- (axis cs:(0.001,9.0) node[anchor=west,rotate=0] { coasttocoastam };
\draw[CB] (axis cs:(0.0003,12.0) node[anchor=west,rotate=0] { anonews\_co };

% CONSPIRACY 

\addplot[mark=triangle,CS,only marks] coordinates {
	(0.8113,102539) %% zerohedge
	(0.13,16435) %% infowars
	(0.0398,5030) %% redflagnews
	(0.0081,1022) %% intellihubnews
	(0.0023,286) %% govtslaves
	(0.0019,246) %% corbettreport
	(0.0016,202) %% realtruthkings
	(0.0016,198) %% hangthebankers
	(0.0009,120) %% beforeitsnews
	(0.0009,113) %% vigilantfeed
};
\draw[CS] (axis cs:(0.91,142539) node[anchor=north,rotate=0] {  {zerohedge} }; %(axis cs:(0.8113,102539)
\draw[CS] (axis cs:(0.13,16435) node[anchor=west,rotate=0] {  {infowars} };
\draw[CS] (axis cs:(0.0398,5030) -- (axis cs:(0.0298,3030) node[anchor=east,rotate=0] {  {redflagnews} };
\draw[CS] (axis cs:(0.0081,1022) node[anchor=east,rotate=0] {  {intellihubnews} };
\draw[CS] (axis cs:(0.0023,286) -- (axis cs:(0.005,300)  node[anchor=west,rotate=0] {  {govtslaves} };
\draw[CS] (axis cs:(0.0019,246) -- (axis cs:(0.005,150) node[anchor=west,rotate=0] {  {corbettreport} };%(axis cs:(0.0015,246) node[anchor=east,rotate=0] { corbettreport };
\draw[CS] (axis cs:(0.0016,202) -- (axis cs:(0.0022,40) node[anchor=west,rotate=0] {  {realtruthkings} };
\draw[CS] (axis cs:(0.0016,198) -- (axis cs:(0.0022,20) node[anchor=west,rotate=0] {  {hangthebankers} };
\draw[CS] (axis cs:(0.0009,120) -- (axis cs:(0.0007,120)node[anchor=east,rotate=0] {  {beforeitsnews} };
\draw[CS] (axis cs:(0.0009,113) --  (axis cs:(0.0008,60) node[anchor=east,rotate=0] {  {vigilantfeed} };

% PROPAGANDA 

\addplot[mark=diamond,P,only marks] coordinates {
	(0.815,496700) %% wikileaks
	(0.0707,43070) %% megynkelly
	(0.0459,28004) %% russiabeyond
	(0.0247,15033) %% therussophile
	(0.0115,7018) %% truthout
	(0.0104,6318) %% southfronteng
	(0.0086,5214) %% engpravda
	(0.0085,5194) %% vdare
	(0.0017,1030) %% regated
	(0.0006,390) %% katehonnews
};
\draw[P] (axis cs:(0.9,496700) node[anchor=north,rotate=0] {  {wikileaks} };%(axis cs:(0.815,496700)
\draw[P] (axis cs:(0.07,40070) node[anchor=west,rotate=0] {  {megynkelly} };% (axis cs:(0.0707,43070)
\draw[P] (axis cs:(0.0459,28004) -- (axis cs:(0.055,8004) node[anchor=west,rotate=0] {  {russiabeyond} }; %(axis cs:(0.045,27004)--(axis cs:(0.055,7504) 
\draw[P] (axis cs:(0.0247,18033) node[anchor=north,rotate=0] {  {therussophile} };% (axis cs:(0.0247,15033) -- (axis cs:(0.035,10033) 
\draw[P] (axis cs:(0.0115,7018) -- (axis cs:(0.001,25018) node[anchor=east,rotate=0] {  {truthout} };%-- (axis cs:(0.003,18018) 
\draw[P] (axis cs:(0.0104,6318) -- (axis cs:(0.001,11018) node[anchor=east,rotate=0] {  {southfronteng} };% -- (axis cs:(0.004,9018) 
\draw[P] (axis cs:(0.0086,5214)  -- (axis cs:(0.003,5000) node[anchor=east,rotate=0] {  {engpravda} };%-- (axis cs:(0.006,5000) 
\draw[P] (axis cs:(0.0085,5194) -- (axis cs:(0.003,2500) node[anchor=east,rotate=0] {  {vdare} };%-- (axis cs:(0.0065,2500)
\draw[P] (axis cs:(0.0017,1030) node[anchor=east,rotate=0] {  {regated} };
\draw[P] (axis cs:(0.0007,750) node[anchor=east,rotate=0] {  {katehonnews} }; %(axis cs:(0.0006,390)

% DISINFO 

\addplot[mark=pentagon,D,only marks] coordinates {
	(0.2272,792987) %% lifenews_ru
	(0.14,488635) %% zvezdanews
	(0.1009,351955) %% wordpressdotcom
	(0.0828,289120) %% vesti_news
	(0.0768,267877) %% lentaruofficial
	(0.0718,250422) %% ntvru
	(0.0532,185806) %% de_sputnik
	(0.0489,170489) %% mfa_russia
	(0.0442,154134) %% novaya_gazeta
	(0.0383,133708) %% topwar_ru
};
\draw[D] (axis cs:(0.2272,792987) -- (axis cs:(0.3,792987) node[anchor=west,rotate=0] { lifenews\_ru };
\draw[D] (axis cs:(0.14,488635) -- (axis cs:(0.25,1608635) node[anchor=west,rotate=0] { zvezdanews };
\draw[D] (axis cs:(0.1009,351955) -- (axis cs:(0.2,3000243) node[anchor=west,rotate=0] { wordpressdotcom };%-- (axis cs:(0.15,351955) 
%\draw[D] (axis cs:(0.0828,289120) -- (axis cs:(0.2,189120) node[anchor=west,rotate=0] { vesti\_news };
%\draw[D] (axis cs:(0.0768,267877) -- (axis cs:(0.18,90877) node[anchor=west,rotate=0] { lentaruofficial };
%\draw[D] (axis cs:(0.0718,250422) --  (axis cs:(0.2,45000) node[anchor=west,rotate=0] { ntvru };%(axis cs:(0.0718,150000) node[anchor=west,rotate=0] { ntvru };
\draw[D] (axis cs:(0.085,299120) -- (axis cs:(0.2,189120) node[anchor=west,rotate=0] { vesti\_news };
\draw[D] (axis cs:(0.0768,267877) -- (axis cs:(0.18,95877) node[anchor=west,rotate=0] { lentaruofficial };
\draw[D] (axis cs:(0.0718,250422) --  (axis cs:(0.25,45000) node[anchor=west,rotate=0] { ntvru };%(axis cs:(0.0718,150000) node[anchor=west,rotate=0] { ntvru };
 
\draw[D] (axis cs:(0.0532,185806) -- (axis cs:(0.08,120000) node[anchor=north,rotate=0]{};% { de\_sputnik };%185806
\draw[D] (axis cs:(0.08,165000) node[anchor=north,rotate=0] { de\_sputnik };%185806 
\draw[D] (axis cs:(0.0489,170489) -- (axis cs:(0.04,35000)%90489)
node[anchor=east,rotate=0] { mfa\_russia };
\draw[D] (axis cs:(0.0442,154134) -- (axis cs:(0.035,60070)%(0.034,45134)
node[anchor=east,rotate=0] { novaya\_gazeta } ;
\draw[D] (axis cs:(0.0383,133708) -- (axis cs:(0.03,113708)  node[anchor=east,rotate=0] {  topwar\_ru  }; %-- (axis cs:(0.02,120708)

%\draw[D] (axis cs:(0.0532,160000) -- (axis cs:(0.035,23000) node[anchor=east,rotate=0] { de\_sputnik };%185806
%\draw[D] (axis cs:(0.0489,170489) -- (axis cs:(0.035,45000) node[anchor=east,rotate=0] { mfa\_russia };
%\draw[D] (axis cs:(0.0442,154134) -- (axis cs:(0.0375,80070) node[anchor=east,rotate=0] { novaya\_gazeta } ;
%\draw[D] (axis cs:(0.0383,133708)  node[anchor=east,rotate=0] {  topwar\_ru  }; %-- (axis cs:(0.02,120708)

%.15	10,20,40,80,160
\addplot[mark=o,V,%green!50!black,
only marks] coordinates { 
	(.23,240) %% regated 
};
\draw[V%green!50!black
] (axis cs:(.23,240)   node[mark=o,anchor=west,rotate=0] {\small Trusted};

\addplot[mark=square,CB,%blue,
only marks] coordinates { 
	(.23,120) %% regated 
};
\draw[CB%blue
] (axis cs:(.23,120)   node[anchor=west,rotate=0] {\small Clickbait};

\addplot[mark=triangle,CS,%orange,
only marks] coordinates { 
	(.23,60) %% regated 
};
\draw[CS%orange
] (axis cs:(.23,60)   node[anchor=west,rotate=0] {\small Conspiracy Theories};

\addplot[mark=diamond,P,%red,
only marks] coordinates { 
	(.23,30) %% regated 
};
\draw[P%red
] (axis cs:(.23,30)   node[anchor=west,rotate=0] {\small Propaganda};

\addplot[mark=pentagon,D,%black,
only marks] coordinates { 
	(.23,15) %% regated 
};
\draw[D%black
] (axis cs:(.23,15)   node[anchor=west,rotate=0] {\small Disinformation};

%\legend{Verified, Clickbait, Conspiracy Theories, Propaganda, Disinformation}

\end{axis}
\end{tikzpicture}
%\end{document}   
		\reduceSpaceBeforeCaption
		\caption{The number of tweets of the ten most frequently occurring news sources within each type as a function of their proportion of all tweets of the given type. For example, anonews\_co is the 10th most frequently occurring clickbait source, but is responsible for only 0.03 \% of the clickbait tweets. In contrast, the most frequent clickbait source theblaze is responsible for 43\%. The 10th (ZeeNews) and most frequent (el\_pais) trusted sources have a much smaller range in terms of shares of the trusted dataset with ZeeNews responsible for 3.5\% and el\_pais for 11.7\%.   }
		\label{fig:top10sources}
	\end{figure*}
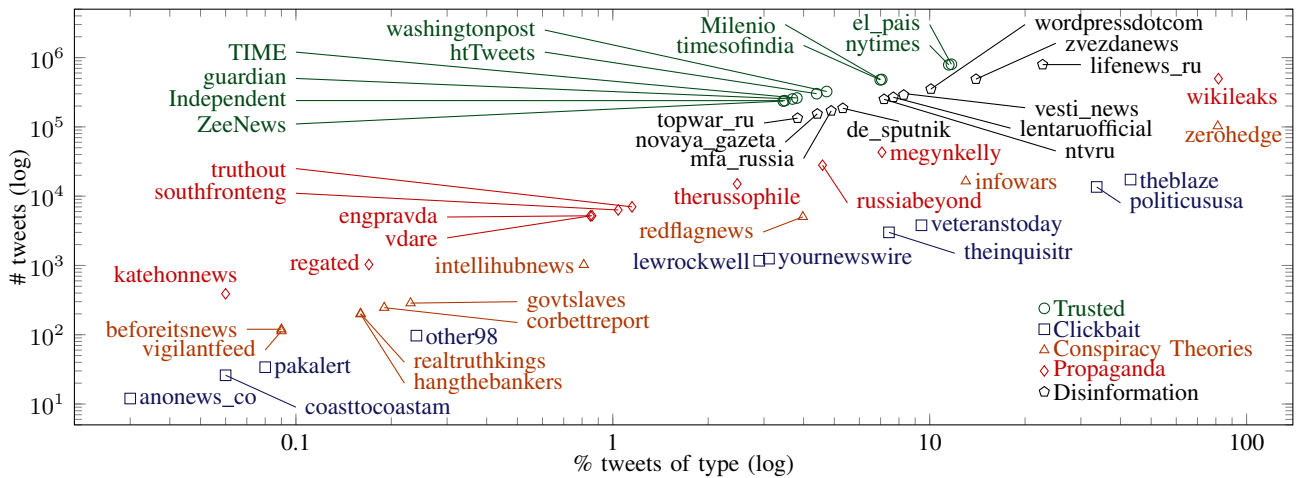

	\vspace{.2in}
	
	\medskip
	\noindent\textbf{\rqfour}  
	
	Existing work focuses on user responses to rumor diffusion as belief exchange that is caused by influence from friends, \eg the Tipping Model~\cite{schelling2006micromotives}. 
	Wu et al. \cite{wu2015false} combined content analysis of rumor-tweets to detect false rumors on Weibo as early as one day after the initial broadcast with 90\% confidence. Among their most important features was the type of user performing the sharing. 
	Ferrara \cite{ferrara2017contagion} used  a similar classification and found those with high followings generated highly-infectious cascades. 
	Studies have also identified that someone who believes in one conspiracy theory is also likely to believe in others~\cite{goertzel1994belief,lewandowsky2013nasa}. Goertzel~\cite{goertzel1994belief} found that "young people were slightly more likely to believe in conspiracy theories" but belief was not significantly correlated with gender or the level of education of the participants. Through our analysis of a sample of users with inferred demographics, we can identify whether there are different patterns in how users interact with conspiracy \textit{sources}.
	In the current work, we focus on a broad question about user sharing behavior to {discover informative patterns within sub-populations not only in the propagation of rumor versus non-rumor or a single category of deceptive content but within interactions spreading information from varying types of news-providers.}

	In order to tackle mis- and disinformation spread in social networks, it is important to address motivations about why people share deceptive news. Motivating factors can be psychological (clickbait), political (propaganda, conspiracy, disinformation), financial (clickbait, disinformation), and social (conspiracy, propaganda) among others. {By analyzing how information from different types of deceptive news sources is propagated across a social network, this study  quantifies how people share from news sources who spread misleading, manipulated, or fabricated information; who these disinformation propagators are; and how much deceptive information is being shared.}

	\section{Data Collection and Annotation}
	In this section we describe how the news sources, the population data, and the demographics data used in this study were collected and annotated with source type labels. 
	\label{data_section}
    \subsection{News Source Labels} 
    As previously discussed, we focus on news sources that fall into one of five classes along a spectrum of varying credibility levels. We define trusted news sources and each of the sub-categories of deceptive news sources as follows:  
 
     \medskip
     \noindent \textbf{Trusted} news sources provide factual information with no intent to deceive the audience, \eg ``\emph{Umberto Eco, Italian semiotician and best-selling author, dies at 84 \emph{\small [URL]} \emph{\small [URL]}}''. 
 
    \medskip
    \noindent \textbf{Clickbait} news sources use attention-grabbing, misleading, or vague headlines such as ``\emph{That's about as tone deaf as it gets right there. \emph{\small [URL]}}'' to attract an audience.
    
    \medskip
    \noindent \textbf{Conspiracy theory} news sources provide uncorroborated or unreliable information to explain events or circumstances. For example, ``\emph{Video: Hoboken train wreck planned? \emph{\small [URL]}}''.
    
    \medskip
    \noindent \textbf{Propaganda} news sources provide intentionally misleading information to advance a social or political agenda, \eg ``\emph{The evidence clearly shows that building new \#nuclear power plants will make global warming worse. \emph{\small [URL]}}''.
    
    \medskip
    \noindent \textbf{Disinformation} news sources share fabricated and factually incorrect information meant to deceive an audience. For example, ``\emph{The great cholesterol and statins con finally unravels: \emph{\small [URL]} \#statins \#cholesterol \#heartdisease \emph{\small [URL]}}''.

	Lists of news sources and their labels were previously aggregated by Volkova et al.~\cite{volkova2017separating} through a combination of crowd-sourcing and public resources. Authors manually constructed a list of trusted news sources that were confirmed to have Twitter-verified accounts that posted content in English. Deceptive news sources, \ie clickbait, conspiracy, and propaganda, were collected from several public resources that annotate suspicious news accounts and their associated websites.\footnote{Deceptive news lists used by Volkova et al~\cite{volkova2017separating} include: \url{http://www.fakenewswatch.com/},  \url{http://www.propornot.com/p/the-list.html}.} Labels for each of the sub-categories of deceptive news sources were also manually verified to ensure quality.  
	
	Disinformation labels were collected from a unique source of public data that comprises confirmed cases of disinformation campaigns: https://euvsdisinfo.eu/, which is also available on Twitter through the {\it @EUvsDisinfo} account. Weekly reports contain disinformation summaries with the countries and languages targeted, as well as the URLs of sources of disinformation, people and organizations who reported disinformation, and manually generated disproofs (when applicable). We limit our analysis to disinformation news accounts collected by the European Union's East Strategic Communications Task Force between 2015 and 2016. As of November 2016, EUvsDisinfo reports include almost 1,992 confirmed disinformation campaigns found in news reports from around Europe and beyond.

	\subsection{Population Data}  
	Our dataset, summarized in Table~\ref{tab:data_summary}, includes approximately 11 million tweets that either retweeted or @mentioned news sources of varying degrees of credibility. We collected all direct mentions of 282 sources over 13 months between January 2016 and January 2017. Then, we assigned the category-label from each news source mentioned or retweeted to each individual tweet. 
	 Figure~\ref{fig:top10sources} shows the relative size and frequency of the ten most frequently occurring sources for each news type in the dataset.

	\begin{table}[t]
		\caption{Summary of our Twitter dataset: the number of news sources, tweets, retweeting (RT) users, and @mentioned users for each news-type and in total over the 13 months between 01/2016 -- 01/2017.}
		\centering
		\small
		\reduceSpaceAroundFigure
		\begin{tabular}{l|r@{\hskip6pt}r@{\hskip6pt}r@{\hskip6pt}r}
			&  Sources &  Tweets &  RT users &  @ users \\
			\hline 
			Trusted (T) &      182 &   6,567,002  &   1,423,227  &           390,164  \\
			Clickbait (CB) &       11  &     40,347 &     19,361  &             6,002 \\
			Conspiracy (CS) &       13  &    126,246 &     35,799  &             9,171  \\
			Propaganda (P) &       26  &    609,251  &    233,799  &            34,532  \\
			Disinformation (D) &       50  &   3,487,732  &    292,437  &            82,638  \\
			\hline
			\bf Total &      \bf 282  &  \bf 10,819,357  &   \bf 1,784,655  &          \bf 471,967  \\ 
		\end{tabular}
		\label{tab:data_summary}
		\vspace{-0.04in}
	\end{table}

	 	\begin{figure}[!ht] 
		\centering 
		\small 
		\hspace{.1in}
\begin{tikzpicture}[trim axis left, trim axis right]
\begin{axis}[ 
height=1in, width=3.25in,
axis lines =left,
legend style={at={(0.5,-0.5)},%at={(1.2,.8)},%
	anchor=north,legend columns=3,column sep=.1cm,draw=none, font=\small},
ylabel={\% retweets},%$P(x)$},  
ylabel style={yshift=-.1in},
ymin=0,ymax=0.35,%1,
ytick = {0,.1,.2,.3},
yticklabels = {0,10,20,30},
symbolic x coords={ 2016-01 , 2016-02 , 2016-03 , 2016-04 , 2016-05 , 2016-06 , 2016-07 , 2016-08 , 2016-09 , 2016-10 , 2016-11 , 2016-12 , 2017-01 ,},
xtick={ 2016-01 , 2016-02 , 2016-03 , 2016-04 , 2016-05 , 2016-06 , 2016-07 , 2016-08 , 2016-09 , 2016-10 , 2016-11 , 2016-12 , 2017-01 ,}, 
xticklabels={ , , , , ,, , , , , , , ,}, 
]
%\node[anchor=north east] at (rel axis cs:0.95,0.85) { \textbf{T} };
\node[anchor=west] at (rel axis cs:0.05,0.8) { \textbf{Trusted} };
 
 %verified
 \addplot[no marks,V, thick] coordinates{
 	 (2016-01 , 8.96327858629e-05 )
 	 (2016-02 , 0.0817722533098 )
 	 (2016-03 , 0.0874472932942 )
 	 (2016-04 , 0.0800563839857 )
 	 (2016-05 , 0.0850115880302 )
 	 (2016-06 , 0.0920696589808 )
 	 (2016-07 , 0.12341062385 )
 	 (2016-08 , 0.100100406238 )
 	 (2016-09 , 0.0861927262556 )
 	 (2016-10 , 0.0859635349302 )
 	 (2016-11 , 0.0867944629711 )
 	 (2016-12 , 0.0721046128477 )
 	 (2017-01 , 0.0189868225207 )
 };

\end{axis}
\end{tikzpicture}

\hspace{.1in}
\begin{tikzpicture}[trim axis left, trim axis right]
\begin{axis}[ 
height=1in, width=3.25in,
axis lines =left,
legend style={at={(0.5,-0.5)},%at={(1.2,.8)},%
	anchor=north,legend columns=3,column sep=.1cm,draw=none, font=\small},
ylabel={\% retweets},%$P(x)$},  
ylabel style={yshift=-.1in},
ymin=0,ymax=0.35,%1,
ytick = {0,.1,.2,.3},
yticklabels = {0,10,20,30},
symbolic x coords={ 2016-01 , 2016-02 , 2016-03 , 2016-04 , 2016-05 , 2016-06 , 2016-07 , 2016-08 , 2016-09 , 2016-10 , 2016-11 , 2016-12 , 2017-01 ,},
xtick={ 2016-01 , 2016-02 , 2016-03 , 2016-04 , 2016-05 , 2016-06 , 2016-07 , 2016-08 , 2016-09 , 2016-10 , 2016-11 , 2016-12 , 2017-01 ,}, 
xticklabels={ , , , , ,, , , , , , , ,}, 
]
%\node[anchor=north east] at (rel axis cs:0.95,0.85) { \textbf{CB} };
\node[anchor=west] at (rel axis cs:0.05,0.8)  { \textbf{Clickbait} };
% clickbait  
\addplot[no marks,CB,loosely dashdotdotted, ultra thick] coordinates {
	 (2016-02 , 0.0703010779333 )
	 (2016-03 , 0.0584066410606 )
	 (2016-04 , 0.0472803865692 )
	 (2016-05 , 0.0423243712056 )
	 (2016-06 , 0.0451988601165 )
	 (2016-07 , 0.258654441829 )
	 (2016-08 , 0.106380869781 )
	 (2016-09 , 0.0629413951183 )
	 (2016-10 , 0.0773881799034 )
	 (2016-11 , 0.083434518647 )
	 (2016-12 , 0.104522364019 )
	 (2017-01 , 0.0431668938174 )
};

\end{axis}
\end{tikzpicture}

\hspace{.1in}
\begin{tikzpicture}[trim axis left, trim axis right]
\begin{axis}[ 
height=1in, width=3.25in,
axis lines =left,
legend style={at={(0.5,-0.5)},%at={(1.2,.8)},%
	anchor=north,legend columns=3,column sep=.1cm,draw=none, font=\small},
ylabel={\% retweets},%$P(x)$},  
ylabel style={yshift=-.1in},
ymin=0,ymax=0.35,%1,
ytick = {0,.1,.2,.3},
yticklabels = {0,10,20,30},
symbolic x coords={ 2016-01 , 2016-02 , 2016-03 , 2016-04 , 2016-05 , 2016-06 , 2016-07 , 2016-08 , 2016-09 , 2016-10 , 2016-11 , 2016-12 , 2017-01 ,},
xtick={ 2016-01 , 2016-02 , 2016-03 , 2016-04 , 2016-05 , 2016-06 , 2016-07 , 2016-08 , 2016-09 , 2016-10 , 2016-11 , 2016-12 , 2017-01 ,}, 
xticklabels={ , , , , ,, , , , , , , ,}, 
]
%\node[anchor=north east] at (rel axis cs:0.95,0.85) { \textbf{CS} };
\node[anchor=west] at (rel axis cs:0.05,0.8) { \textbf{Conspiracy} };
% conspiracy  
\addplot[no marks,CS,dashed,ultra thick] coordinates {
	 (2016-01 , 0)
	 (2016-02 , 0.0687977212486 )
	 (2016-03 , 0.056114254065 )
	 (2016-04 , 0.0492305257744 )
	 (2016-05 , 0.0570241721723 )
	 (2016-06 , 0.0817976816869 )
	 (2016-07 , 0.206401076077 )
	 (2016-08 , 0.0746449341298 )
	 (2016-09 , 0.0638999881315 )
	 (2016-10 , 0.0896229774103 )
	 (2016-11 , 0.0968785852752 )
	 (2016-12 , 0.108667959014 )
	 (2017-01 , 0.0469201250148 )
};

\end{axis}
\end{tikzpicture}

\hspace{.1in}
\begin{tikzpicture}[trim axis left, trim axis right]
\begin{axis}[ 
height=1in, width=3.25in,
axis lines =left,
legend style={at={(0.5,-0.5)},%at={(1.2,.8)},%
	anchor=north,legend columns=3,column sep=.1cm,draw=none, font=\small},
ylabel={\% retweets},%$P(x)$},  
ylabel style={yshift=-.1in},
ymin=0,ymax=0.35,%1,
ytick = {0,.1,.2,.3},
yticklabels = {0,10,20,30},
symbolic x coords={ 2016-01 , 2016-02 , 2016-03 , 2016-04 , 2016-05 , 2016-06 , 2016-07 , 2016-08 , 2016-09 , 2016-10 , 2016-11 , 2016-12 , 2017-01 ,},
xtick={ 2016-01 , 2016-02 , 2016-03 , 2016-04 , 2016-05 , 2016-06 , 2016-07 , 2016-08 , 2016-09 , 2016-10 , 2016-11 , 2016-12 , 2017-01 ,}, 
xticklabels={ , , , , ,, , , , , , , ,}, 
]
%\node[anchor=north east] at (rel axis cs:0.95,0.85) { \textbf{P} };
\node[anchor=west] at (rel axis cs:0.05,0.8) { \textbf{Propaganda} };
 
% propaganda 
\addplot[no marks,P, loosely dashdotted, ultra thick] coordinates {
	 (2016-01 , 3.28161507968e-06 )
	 (2016-02 , 0.0380306371584 )
	 (2016-03 , 0.0292326271298 )
	 (2016-04 , 0.0364259273844 )
	 (2016-05 , 0.0293868630385 )
	 (2016-06 , 0.0248730014964 )
	 (2016-07 , 0.167199929117 )
	 (2016-08 , 0.0607049565514 )
	 (2016-09 , 0.0319120658423 )
	 (2016-10 , 0.297548961697 )
	 (2016-11 , 0.186448242367 )
	 (2016-12 , 0.0752441521619 )
	 (2017-01 , 0.0229893544407 )
};

\end{axis}
\end{tikzpicture}

%\vspace{-.03in}
%\vspace{-.07in}
%\hspace{.2in}
\hspace{.15in}%74in}
\begin{tikzpicture}[trim axis left, trim axis right]
\begin{axis}[ 
height=1in, width=3.25in,
axis lines =left,
legend style={at={(0.5,-0.5)},%at={(1.2,.8)},%
	anchor=north,legend columns=3,column sep=.1cm,draw=none, font=\small},
xlabel={Month Retweeted}, xlabel style={yshift=-.25in},
ylabel={\% retweets},%$P(x)$},  
ylabel style={yshift=-.1in},
ymin=0,ymax=0.35,%1,
ytick = {0,.1,.2,.3},
yticklabels = {0,10,20,30},
symbolic x coords={ 2016-01 , 2016-02 , 2016-03 , 2016-04 , 2016-05 , 2016-06 , 2016-07 , 2016-08 , 2016-09 , 2016-10 , 2016-11 , 2016-12 , 2017-01 ,},
xtick={ 2016-01 , 2016-02 , 2016-03 , 2016-04 , 2016-05 , 2016-06 , 2016-07 , 2016-08 , 2016-09 , 2016-10 , 2016-11 , 2016-12 , 2017-01 ,}, 
xticklabels={Jan 2016, ,Mar 2016, ,May 2016,,July 2016, ,Sept 2016, ,Nov 2016, ,Jan 2017,}, %{,Feb 2016, ,Apr 2016, ,Jun 2016,,Aug 2016, ,Oct 2016, ,Dec 2016, ,}, %
xticklabel style ={anchor=east,yshift=-1mm,
	xshift=3mm,rotate=30}, 
]
%\node[anchor=north east] at (rel axis cs:0.95,0.85) { \textbf{D} };
\node[anchor=west] at (rel axis cs:0.05,0.8)  { \textbf{Disinformation} };

% disinfo
\addplot[no marks,D,densely dotted, ultra thick] coordinates {
 (2016-01 , 1.1461915779e-06 )
 (2016-02 , 0.0931753461069 )
 (2016-03 , 0.0961193391747 )
 (2016-04 , 0.0880616123821 )
 (2016-05 , 0.0835502023315 )
 (2016-06 , 0.0809698385417 )
 (2016-07 , 0.0922211416183 )
 (2016-08 , 0.0954791911785 )
 (2016-09 , 0.08807823216 )
 (2016-10 , 0.0893227096657 )
 (2016-11 , 0.0841940754504 )
 (2016-12 , 0.0825306649229 )
 (2017-01 , 0.0262965002759 )
};

\end{axis}
\end{tikzpicture}  
		\reduceSpaceBeforeCaption
		\caption{
		Tweet volume over the 13 months between 01/2016 -- 01/2017. 
		The percentage of tweets within each source-type is plotted as a function of the month posted or shared.}
		\label{fig:month_shared} 
	\end{figure}
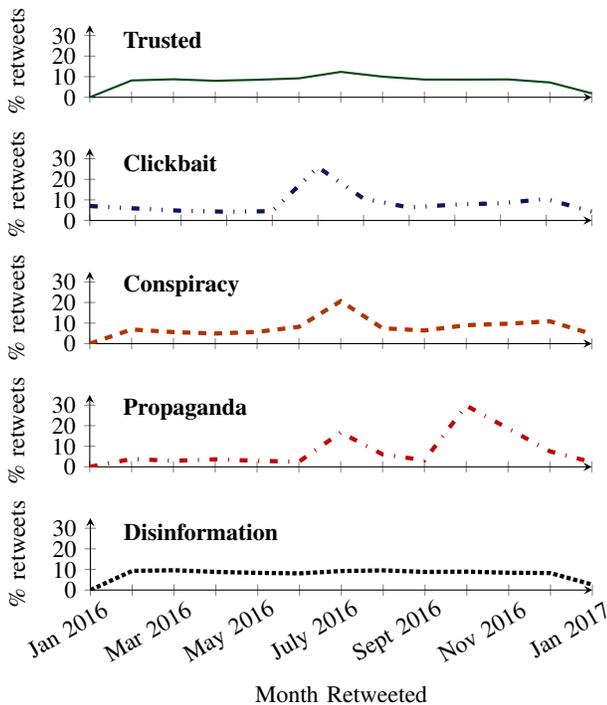

	Distributions of tweet volume over time are illustrated in Figure~\ref{fig:month_shared}. Trusted and disinformation-spreading sources are referenced and retweeted consistently across the entire year. The other three deceptive types have concentrated peaks in activity sharing or mentioning source accounts. Clickbait- and Conspiracy-spreading sources both have a single peak in June 2016 (26\%) and July 2016 (21\%), respectively. Tweets that reference or retweet propaganda-spreading sources were most heavily posted in October and November 2016 (49\%), with another spike in July 2016 (17\%).

	Recent work has found that accounts spreading disinformation are significantly more likely to be automated accounts~\cite{shao2017spread}.  
	However, in the current work we are interested in the impact that news sources have in the system as a whole, capturing the publicly visible responses to news sources by all accounts ---  whether activity from that account is manual, automated, or some mixture of both manual and automated behavior. Therefore, we did not remove bots or automated accounts from the population dataset. For the dataset we used in our demographics-based analysis, described in the next subsection, we focus specifically on personal accounts. 
	  
	 We did compare the news accounts and tweets captured in our dataset with the list of Twitter accounts connected to Russia's ``Internet Research Agency'' recently released by the US House Intelligence committee.\footnote{\url{https://democrats-intelligence.house.gov/uploadedfiles/exhibit_b.pdf}} However, because we are focused primarily on sources of news (rather than specific rumors or events), our dataset only contained 180 of the 2752 flagged-accounts, and only 4763 retweets; so we did not focus on these accounts in particular in the current work.

	\subsection{Demographics Data}
	%\subsection{User Level} 
	In addition to the population-level aggregate statistics, we %are the first to 
	study information propagation and influence across various user demographics. To accurately obtain user demographics we curated a subset of non-organization user accounts that generated enough information to render a demographic description with high confidence. To identify this sample, we first restrict the dataset to users who 1) retweeted or @mentioned deceptive news sources in at least 5 posts during the data collection period, and 2) posted in English. This resulted in a subset of 106,849 users. From this list we collected the 200 most recent tweets from Twitter's public API. 
	We used these tweets and the Humanizr classifier to identify person-accounts, which they defined as "a personal account is one controlled by an individual"~\cite{mccorriston2015organizations}.  This resulted in a sub-sample of {\bf 66,171 person accounts}.% which are the focus of our demographics-based analysis.
	
	To infer user demographics, \ie {\bf gender, age, income, and education}, we employed a demographic classifier trained on a large, previously annotated Twitter dataset~\cite{volkova2016inferring}. Specifically, our demographic classifier relied on a Convolutional Neural Network (CNN) architecture initialized with 200-dimensional GLOVE embedding vectors pre-trained on Twitter tokens~\cite{pennington2014glove}. Following previous methodology, each demographic-attribute was assigned one of two mutually exclusive classes~\cite{volkova2016inferring}. For example, we classified gender as either male (M) or female (F), age as either younger than 25 (Y) or 25 and older (O), income as below (B) or at and above (A) \$35,000 a year, and education as having only a high school education (H) or at least some college education (C). The area under the ROC curve (AUC) for 10-fold cross-validation experiments were {\bf 0.89 for gender, 0.72 for age, 0.72 for income, and 0.76 for education}. These are the state-of-the-art results for user demographics prediction on Twitter, an improvement on performance of previous models that used the same dataset~\cite{volkova2015predicting}. 
	
		Users in this dataset were primarily predicted to be male (96\%), older (95\%), with higher incomes (81\%), college-educated (82\%), and classified as regular users (59\%).
	It is important to note that our user sample is representative of those users who frequently interact with deceptive source accounts.  
	It is not a balanced sample of global demography or a representative sample of Twitter itself. In fact, a survey by the Pew Research Center in 2016 found that only 17\% of Twitter users had a high school education or less, 38\% were between 18 and 29 years old, and 47\% were male~\cite{gottfried2016pew}. Each reported category was less skewed towards the majority class in our demographic attributes than we found in the sample of users who frequently interacted with deceptive source accounts.% in our dataset.
	%
	% Pew Research Center news use 2016 survey
	% age: 38\% 18 - 29, 61\% 30+
	% gender: 47\% male, 53\% female
	% education: 45\% college degree, 38\% some college, 17\% high school or less
	% 

	We also include each user's role in their network based on the friend and follower counts collected from user metadata. For this analysis we borrow the leader/follower heuristic to assign a user as an {\bf opinion leader (L)} if they are followed by more users than they follow or a {\bf regular user (R)} if they follow more users than they have followers~\cite{wu2015false}.

	\section{Methodology}
	In this section, we describe the methodology used to analyze propagation behavior of news content and misinformation across and within the five types of sources identified. Again, we focus on propagation at the \textit{source} level rather than the content or individual tweet level. We consider propagation of all content, deceptive or not, from sources of each type.
	
	% RQ1 : How evenly
	
	\rqonesection To answer this question we compare the distributions of how users share information using three measures commonly used to measure income inequality: Lorenz curves, Gini coefficients, and Palma ratios. Rather than measure how much of the total population's income each individual is responsible for, we repurpose these metrics to measure how much of the total tweet volume each user is responsible for. This allows us to compare \textit{propagation inequality} across source types the way economists compare income inequality across countries.% or regions. 
	
\begin{figure}[ht]%[!ht]
		\centering
		\small
		\usepgfplotslibrary{fillbetween}
\begin{tikzpicture}
\begin{axis}[
axis lines = left,
clip=false,
height=2.25in, width=2.2in,
xlabel style={yshift=0.05in,align=center}, 
ylabel style={yshift=-0.1in,align=center},
legend cell align={left},
legend style={at={(.5,-.25)}, anchor=north,legend columns=-1,draw=none,fill=none,font=\small},
xlabel={cumulative \% active users \\ \textcolor{gray!50!black}{(cumulative \% population)}},
ylabel={\% propagation (tweets shared)\\ \textcolor{gray!50!black}{(\% income)}},
xmin=0-.1,xmax=1.1,ymin=0-.1,ymax=1.1,
xtick = {0,.25,.5,.75,1},xticklabels ={0,25,50,75,100},
ytick = {0,.25,.5,.75,1},yticklabels ={0,25,50,75,100}, 
xtick align=center, 
ytick align=center, 
title ={Propagation Inequality\\ \textcolor{gray!50!black}{(Income Inequality)}}, title style = {yshift= -.1in, xshift=-.1in,align=center},
] 

%Perfect Equality
\addplot[black,no marks,thick, name path=perfectEquality] coordinates {
	(0,0) (1,1)
};  

%\draw[black] (axis cs:(0.5,0.5) node[anchor=south,rotate=45] {  {\tiny Lorenz Curve of Perfect Equality} };
 
%Perfect Inequality
\addplot[black,no marks,very thick,densely dotted] coordinates {
	(0.001,0.001) 
	(1,0.001) (1,1)
};

\draw[black!50] (75,55) node[color=black] {\normalsize $a_1$} ;
\draw[black!50] (95,20) node[anchor=south,color=black] {\normalsize $a_2$} ;

% propaganda with mentions
\addplot[no marks,black, dashed, thick, name path=lorenz] coordinates {
	(1,1)
	% (2016+) propaganda, Gini:0.527386049892, Palma:3.13816751094
	( 1 - 0.00999947940235 , 1 - 0.251670554094 )  ( 1 - 0.0199989588047 , 1 - 0.311754693227 )  ( 1 - 0.0299984382071 , 1 - 0.353967063288 )  ( 1 - 0.0399979176094 , 1 - 0.387767161445 )  ( 1 - 0.0499973970118 , 1 - 0.415830500518 )  ( 1 - 0.0599968764141 , 1 - 0.4404222386 )  ( 1 - 0.0699963558165 , 1 - 0.461292880017 )  ( 1 - 0.0799998398161 , 1 - 0.481728545424 )  ( 1 - 0.0899993192185 , 1 - 0.498123037478 )  ( 1 - 0.0999987986208 , 1 - 0.514517529533 )  ( 1 - 0.109998278023 , 1 - 0.530912021588 )  ( 1 - 0.119997757426 , 1 - 0.543518118802 )  ( 1 - 0.129997236828 , 1 - 0.555813987843 )  ( 1 - 0.13999671623 , 1 - 0.568109856885 )  ( 1 - 0.149996195633 , 1 - 0.580405725926 )  ( 1 - 0.159999679632 , 1 - 0.592706519223 )  ( 1 - 0.169999159035 , 1 - 0.604337613607 )  ( 1 - 0.179998638437 , 1 - 0.612534859634 )  ( 1 - 0.189998117839 , 1 - 0.620732105662 )  ( 1 - 0.199997597242 , 1 - 0.628929351689 )  ( 1 - 0.209997076644 , 1 - 0.637126597717 )  ( 1 - 0.219996556046 , 1 - 0.645323843744 )  ( 1 - 0.229996035449 , 1 - 0.653521089771 )  ( 1 - 0.239999519448 , 1 - 0.661721618637 )  ( 1 - 0.249998998851 , 1 - 0.669918864664 )  ( 1 - 0.259998478253 , 1 - 0.678116110692 )  ( 1 - 0.269997957655 , 1 - 0.686313356719 )  ( 1 - 0.279997437058 , 1 - 0.694510602747 )  ( 1 - 0.28999691646 , 1 - 0.702707848774 )  ( 1 - 0.299996395862 , 1 - 0.710905094802 )  ( 1 - 0.309999879862 , 1 - 0.717180239287 )  ( 1 - 0.319999359264 , 1 - 0.721278862301 )  ( 1 - 0.329998838667 , 1 - 0.725377485315 )  ( 1 - 0.339998318069 , 1 - 0.729476108328 )  ( 1 - 0.349997797471 , 1 - 0.733574731342 )  ( 1 - 0.359997276874 , 1 - 0.737673354355 )  ( 1 - 0.369996756276 , 1 - 0.741771977369 )  ( 1 - 0.379996235679 , 1 - 0.745870600383 )  ( 1 - 0.389999719678 , 1 - 0.749970864815 )  ( 1 - 0.399999199081 , 1 - 0.754069487829 )  ( 1 - 0.409998678483 , 1 - 0.758168110842 )  ( 1 - 0.419998157885 , 1 - 0.762266733856 )  ( 1 - 0.429997637288 , 1 - 0.766365356869 )  ( 1 - 0.43999711669 , 1 - 0.770463979883 )  ( 1 - 0.449996596092 , 1 - 0.774562602897 )  ( 1 - 0.459996075495 , 1 - 0.77866122591 )  ( 1 - 0.469999559494 , 1 - 0.782761490343 )  ( 1 - 0.479999038897 , 1 - 0.786860113356 )  ( 1 - 0.489998518299 , 1 - 0.79095873637 )  ( 1 - 0.499997997701 , 1 - 0.795057359384 )  ( 1 - 0.509997477104 , 1 - 0.799155982397 )  ( 1 - 0.519996956506 , 1 - 0.803254605411 )  ( 1 - 0.529996435908 , 1 - 0.807353228424 )  ( 1 - 0.539999919908 , 1 - 0.811453492857 )  ( 1 - 0.54999939931 , 1 - 0.81555211587 )  ( 1 - 0.559998878713 , 1 - 0.819650738884 )  ( 1 - 0.569998358115 , 1 - 0.823749361898 )  ( 1 - 0.579997837517 , 1 - 0.827847984911 )  ( 1 - 0.58999731692 , 1 - 0.831946607925 )  ( 1 - 0.599996796322 , 1 - 0.836045230938 )  ( 1 - 0.609996275725 , 1 - 0.840143853952 )  ( 1 - 0.619999759724 , 1 - 0.844244118384 )  ( 1 - 0.629999239127 , 1 - 0.848342741398 )  ( 1 - 0.639998718529 , 1 - 0.852441364412 )  ( 1 - 0.649998197931 , 1 - 0.856539987425 )  ( 1 - 0.659997677334 , 1 - 0.860638610439 )  ( 1 - 0.669997156736 , 1 - 0.864737233452 )  ( 1 - 0.679996636138 , 1 - 0.868835856466 )  ( 1 - 0.689996115541 , 1 - 0.87293447948 )  ( 1 - 0.69999959954 , 1 - 0.877034743912 )  ( 1 - 0.709999078943 , 1 - 0.881133366926 )  ( 1 - 0.719998558345 , 1 - 0.885231989939 )  ( 1 - 0.729998037747 , 1 - 0.889330612953 )  ( 1 - 0.73999751715 , 1 - 0.893429235966 )  ( 1 - 0.749996996552 , 1 - 0.89752785898 )  ( 1 - 0.759996475954 , 1 - 0.901626481994 )  ( 1 - 0.769999959954 , 1 - 0.905726746426 )  ( 1 - 0.779999439356 , 1 - 0.90982536944 )  ( 1 - 0.789998918759 , 1 - 0.913923992453 )  ( 1 - 0.799998398161 , 1 - 0.918022615467 )  ( 1 - 0.809997877563 , 1 - 0.922121238481 )  ( 1 - 0.819997356966 , 1 - 0.926219861494 )  ( 1 - 0.829996836368 , 1 - 0.930318484508 )  ( 1 - 0.839996315771 , 1 - 0.934417107521 )  ( 1 - 0.84999979977 , 1 - 0.938517371954 )  ( 1 - 0.859999279172 , 1 - 0.942615994967 )  ( 1 - 0.869998758575 , 1 - 0.946714617981 )  ( 1 - 0.879998237977 , 1 - 0.950813240995 )  ( 1 - 0.88999771738 , 1 - 0.954911864008 )  ( 1 - 0.899997196782 , 1 - 0.959010487022 )  ( 1 - 0.909996676184 , 1 - 0.963109110035 )  ( 1 - 0.919996155587 , 1 - 0.967207733049 )  ( 1 - 0.929999639586 , 1 - 0.971307997481 )  ( 1 - 0.939999118989 , 1 - 0.975406620495 )  ( 1 - 0.949998598391 , 1 - 0.979505243509 )  ( 1 - 0.959998077793 , 1 - 0.983603866522 )  ( 1 - 0.969997557196 , 1 - 0.987702489536 )  ( 1 - 0.979997036598 , 1 - 0.991801112549 )  ( 1 - 0.989996516 , 1 - 0.995899735563 )
	(0,0) 
	
};

\addplot[gray!50] fill between[of=perfectEquality and lorenz];

%\legend{Perfect Equality,Perfect Inequality,Lorenz Curve} 
\end{axis}
\end{tikzpicture}

%\vspace{-.05in}
\begin{tikzpicture} 
\begin{axis}[%
hide axis, height =.75in,
xmin=0,xmax=50,ymin=0,ymax=0.4,
legend style={at={(0.5,1)},%0.5)},
	anchor=north,legend columns=-1,column sep=.05cm,draw=none},   
] 
\addlegendimage{black,solid, thick}
\addlegendentry{Perfect Equality}; 
\addlegendimage{black,densely dotted, thick}
\addlegendentry{Perfect Inequality}; 
%\addlegendimage{black,dashed, thick}
%\addlegendentry{Lorenz Curve}; 
\end{axis}
\end{tikzpicture}

\vspace{-.05in}
\begin{tikzpicture} 
\begin{axis}[%
hide axis, height =.75in,
xmin=0,xmax=50,ymin=0,ymax=0.4,
legend style={at={(0.5,1)},%0.5)},
	anchor=north,legend columns=-1,column sep=.05cm,draw=none,fill=none},   
]  
\addlegendimage{black,dashed, thick}
\addlegendentry{Example Lorenz Curve}; 
\end{axis}
\end{tikzpicture}

		%\reduceSpaceBeforeCaption
		\caption{ Lorenz Curves and Gini Coefficients. 
		As a graphical representation of income inequality within a population, Lorenz curves plot the share of income by the cumulative share of the population. Lorenz curves that measure the inequality in propagation plot the share of total propagation, \ie the $y$\% of tweets posted, by the share of the population who propagated, \ie the cumulative $x$\% of active users. The Gini coefficient is the proportion of the area under the line of perfect equality ($a_1+a_2$) that is captured between the line of perfect equality and the Lorenz curve ($a_1$). (RQ1)}%That is, $G = {\dfrac{a_1}{a_1+a_2}} $. (RQ1)}
		%\vspace{.2in}
		%\reduceSpaceAroundFigure
		\label{fig:lorenz_gini_example}
	\end{figure}
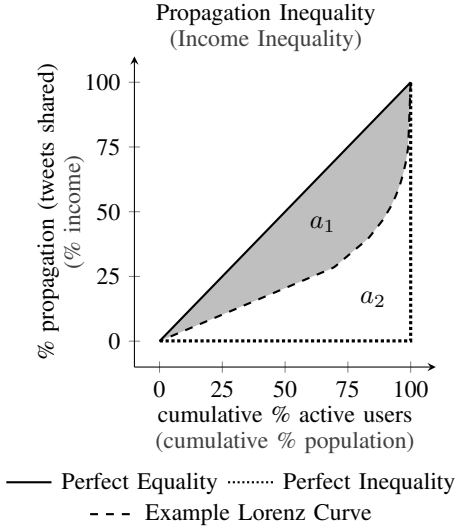

	Lorenz curves have traditionally been used to illustrate the distribution of income or wealth graphically~\cite{kakwani1973estimation}. In those domains, the curves plot the cumulative percentage of wealth, income, or some other variable to be compared against the cumulative (in increasing shares) percentage of a corresponding population. The degree to which the curve deviates from the straight diagonal ($y=x$) representative of perfect equality represents the inequality present in the distribution. In our case, the Lorenz curve is adapted to illustrate the cumulative percentage of propagation (tweets shared) as a function of the cumulative percentage of users posting, as shown in Figure~\ref{fig:lorenz_gini_example}. 
	
	The Gini coefficient is defined as the proportion of the area under the line of perfect equality that is captured above the Lorenz curve, \ie $\frac{a_1}{a_1+a_2}$ in Figure~\ref{fig:lorenz_gini_example}. 
	The Gini coefficients reported in subsequent sections are calculated using the formula in Eq.~\eqref{eq:gini}, which is an approximation of the points of the Lorenz curves observed in the collected data. Using income for our example, Gini coefficients can grow to be greater than 1 but only if individuals within a population can be responsible for negative proportions, \eg if individuals can have negative incomes. In our data, users must be responsible for at least 1 share in order to be considered part of the dataset, so the Gini coefficients have an upper-bound of 1. 
	\begin{equation}
	\hat{G} =  1 - \sum_{k=1}^{n} (X_k - X_{k-1})(Y_k + Y_{k-1})
	\label{eq:gini}
	\end{equation}
	
	The third measure we consider is the Palma ratio.
	It is defined as the ratio of the share of the top 10\% to the bottom 40\% of users in the population. Again, using income as our example, in perfect equality each individual in the population is responsible for an equal amount, \eg an equal share of income, resulting in a Palma ratio of $1/4$.  
	The Palma ratio was formulated as another measure of inequality because the Gini coefficient is most sensitive to changes in the middle, which is relatively stable \cite{Cobham13puttingthe}. The Palma ratio, on the other hand, is sensitive to changes at the extremes. 
	
	\smallskip
	We use the Gini coefficient, Palma ratio, and the Lorenz curves to provide a balanced understanding of the inequalities in the distributions of how information is spread online across the five types of news sources.
	\smallskip

	% RQ2 : how many users
	
	\rqtwosection To answer this question we measure the size of the Twitter audience that directly retweets content as a measure of source influence. We compare the distributions of and the average number of users who propagate content posted by sources for each source type. This allows us to measures how large of a direct response each source post causes across source types. When we consider the results at the user level, we compare the behavior of each sub-population of users.

	% RQ3 : How quickly 
	
	\rqthreesection To answer this question we measure the speed by which users share content with their followers. Specifically, we measure the delay from the original tweet from the news source to the time of the retweet. It is important to note that it is not the goal of the present work to measure the entire cascade of information propagation; rather, we are interested in direct retweets of news source accounts and, therefore, only collect these specific share events. To compensate for these methodological decisions, we borrow similar statistics from recent work that measured all share events~\cite{ferrara2015quantifying}. We can then draw conclusions by comparing our measurements of direct shares against the global aggregate.

	% RQ4: Who is sharing?
	
	\rqfoursection To answer this question at an aggregate level we look at user overlap and user-network similarities across the five types of news sources. We hypothesize that there will be large overlaps with trusted sources for the sets of users interacting with sources identified as spreading deceptive content, but that this overlap will probably not be symmetric. That is, that users who spread content from suspicious sources may also spread content from trusted sources. However, users who spread content from trusted sources may be less likely to also spread content from suspicious sources. 
	
	Then, we compare source types by certain social network statistics including density, edges to nodes ratios, or average indegree, outdegree, shortest path length, etc. Each source is represented as its own social graph (sub-network). Nodes in each graph represent users who retweeted or @mentioned the news source, or who were @mentioned (using @user) in a tweet connected to the source (through an @mention or a retweet of the source by another user). Edges represent the links between these users on a per-source basis. Specifically, we draw an edge between users $x$ and $y$ if $x$ retweets $y$, $x$ includes @$y$ in a tweet, or $z$ mentions both @$x$ and @$y$ in the same tweet. 
	
	We also measure shares across  various user demographics and five news source types. We present the results of Mann Whitney U (MWU) tests to identify statistically significant differences in who is sharing information for each type of news, along with common language effect sizes to illustrate the magnitude of those differences.

	\section{Deception Propagation Results: Population Level}
	We first look at the behavior of users spreading information from each type of news source at the population level. Here we present the results of experiments that use the large dataset of almost 11 million tweets over the course of 2016. We compare and contrast how evenly, how much, how quickly, and who shares content within and across user sub-populations interacting with different types of news sources --- trusted, clickbait, conspiracy, propaganda, and disinformation.

	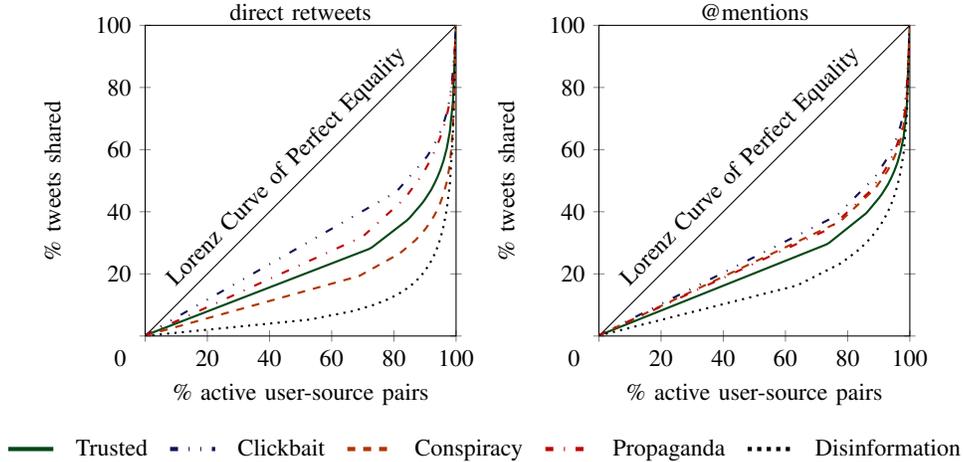
\begin{figure*}[t!] 
		\centering
		\small
		\include{figs/journal/lorenz_curves_population}
		\vspace{-.2in} 
		\begin{tikzpicture} 
\begin{axis}[%
hide axis, height =.75in,
xmin=0,xmax=50,ymin=0,ymax=0.4,
legend style={at={(0.5,1)},%0.5)},
	anchor=north,legend columns=-1,column sep=.2cm,draw=none},   
]
\addlegendimage{no marks, V,ultra thick}
\addlegendentry{Trusted};%Verified};
\addlegendimage{no marks,CB,loosely dashdotdotted, ultra thick}
\addlegendentry{Clickbait};%Clickbait};
\addlegendimage{no marks,CS,dashed, ultra thick}
\addlegendentry{Conspiracy};%Conspiracy}; 
\addlegendimage{no marks,P,loosely dashdotted, ultra thick}
\addlegendentry{Propaganda};%Propaganda};
\addlegendimage{no marks,D,dotted, ultra thick}
\addlegendentry{Disinformation};%Disinformation}; 
\end{axis}
\end{tikzpicture}
		\reduceSpaceBeforeCaption
		\caption{ 
		Lorenz curves of the propagation inequality for each source type, averaged across sources using user-source pairs instead of users as diffusion units. Propagation using direct retweets at left and @mention tweets at right. The lorenz curve representative of perfect equality is also included for reference. (RQ1)}
		%\reduceSpaceAroundFigure
		\label{fig:retweet_diffusion_lorenz_curves}
	\end{figure*} 
	
	\rqonesection Across all types, including trusted sources, we see large diffusion inequality. We find that a relatively small subset of users are responsible for a large proportion of each sources' shares. However, these inequalities are not equal across news source types. To look at information propagation inequality for each source type overall, we use each source-user pair as a single propagation unit in the propagation frequency distributions used to generate Lorenz curves and calculate Gini coefficients and Palma ratios for each source type.  
	
	We illustrate the inequalities in the propagation of each source type with their respective Lorenz curves in Figure~\ref{fig:retweet_diffusion_lorenz_curves}. The curve for perfect equality is also included for easy comparison not only across news types but within the context of best and worst cases. We see that, for direct retweets, the Lorenz curve for disinformation sources is the furthest from the line of perfect equality. There is a significant gap ($p<0.01$) between it and the next closest Lorenz curve (conspiracy sources). Except for conspiracy theory sources which are more equally diffused, the arrangement of Lorenz curves for each source type from closest to furthest from perfect equality is the same for @mentions as for retweets. While Lorenz curves for @mention tweets are more closely plotted around the curve for propaganda sources, there is still significantly more inequality in the propagation of disinformation sources ($p<0.01$).

	Table \ref{tab:gini_palma} presents the Gini coefficients and Palma ratios for each news type. Again, we see the greatest inequality in retweet diffusion for disinformation sources, although much lower for @mentions of those sources than direct retweets. The 10\% most active users who directly retweet disinformation-spreading sources share 20.13 times as many tweets as the least active 40\%, compared to around 2 - 6 times as much for each of the other source types. This ratio drops to 6.49 for @mentions of disinformation sources.

	\begin{table}[!ht] 
		\caption{ 
		Gini coefficients and Palma ratios for direct retweets (RT) from or tweets that @mention sources (@) for each news type, averaged across sources using source-user combinations as diffusion units. Higher values mean more inequality. (RQ1)}
		% in the retweet frequency distribution
		\centering
		\small  
    	%\reduceSpaceAroundFigure
    	\setlength{\tabcolsep}{0.3em}
    	  
    	\begin{tabular}{l|rrr@{\hskip20pt}rr}
    	                & \multicolumn{3}{c}{\sc Gini Coefficient} & \multicolumn{2}{c}{\sc Palma Ratio}\\
    	   Source-Type && RT & @ & RT & @ \\ \hline
    	     Trusted &&  0.57& 0.56  &  3.68& 3.52 \\
    	     Clickbait && 0.40&  0.46&  1.91&  2.37 \\
    	     Conspiracy &&  0.67& 0.49 & 5.95& 2.73 \\
    	     Propaganda &&  0.48& 0.49 &  2.61& 2.71 \\
    	     Disinformation~~ &&  \textbf{0.83}&  0.68 &  \textbf{20.13}&  \textbf{6.49} \\ 
    	\end{tabular}
    	\reduceSpaceAroundFigure

		\label{tab:gini_palma}
	\end{table}
	
	Shares of @mention propagation are more equally distributed among users than direct retweets for trusted, conspiracy, and disinformation. Direct retweet of clickbait and propaganda sources are more evenly shared by users than @mentions, but only slightly. The Gini coefficients also illustrate the larger inequality gap for disinformation sources that we saw with the Lorenz curves, reaching 0.83 for direct retweets. 
	All other source types except for direct retweets of conspiracy sources have Gini coefficients below the minimum coefficient for a Pareto 80:20 distribution (0.60). Interestingly, we see that clickbait and propaganda sources are more evenly propagated than trusted sources. The more-even distribution of clickbait articles is not surprising --- the whole point of clickbait articles is to motivate many people to click and share the articles.

    We also compared the Gini coefficients and Palma ratios of individual sources. The only statistically significant findings ($p<0.01$) were in the differences between disinformation-spreading sources and all other types of news sources. In particular, disinformation sources had higher Gini coefficients than trusted sources in 63\% of comparisons, and propaganda sources in 66\% of comparisons. Using Palma ratios, disinformation sources had higher ratios than trusted sources and propaganda sources in 72\% and 75\% of comparisons, respectively. These results show that the volume of retweets for disinformation sources are more unevenly distributed than trusted or propaganda sources; these results also demonstrate that the unevenness of distribution is more heavily evident in the extremes of the distributions --- among the 10\% most and 40\% least active users.

	\rqtwosection We compared the mean number of users who retweet each source post and found that the 95\% confidence intervals of those means overlap for all source type comparisons except between conspiracy and disinformation sources. This is not unexpected because the retweet distributions are so heavily skewed. Long tails may heavily influence the means. However, we found that the distributions of the number of users who retweet each source post for trusted and disinformation sources differ significantly ($p<0.05$). Further, when we compare the distributions of the mean number of users who retweeted each source tweet for each of the sources, we also see statistically significant differences ($p=0.03$) where the mean number of users who retweeted a disinformation source is greater than that of a trusted source in 60\% of comparisons. Disinformation sources have, on average, more users retweet each source post.

	\begin{figure}[!t]
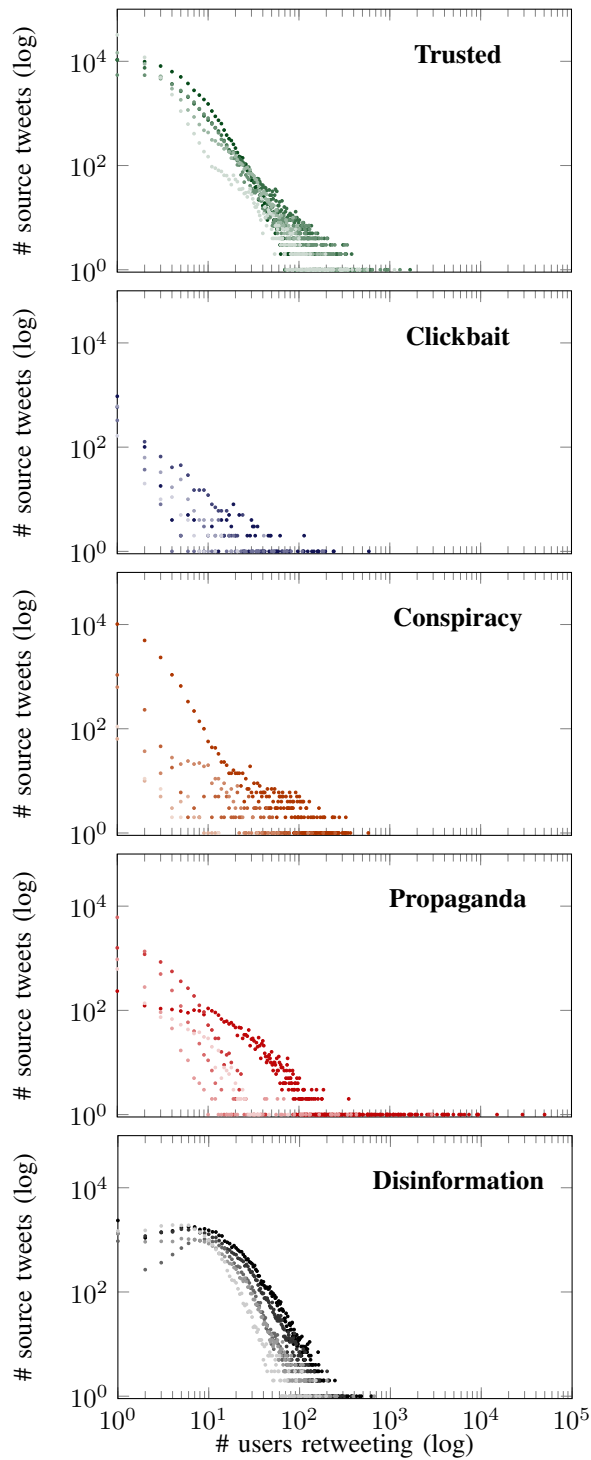

	    %\reduceSpaceAroundFigure
		\centering      
		\include{figs/journal/rt_user_volume_top5_by_label_pop_vertical}
		%\reduceSpaceBeforeCaption
		\caption{ 
		Distributions of the number of source tweets with a given number of users who retweeted for the five most frequently occurring sources in each type. More frequently occurring sources are plotted in darker shades. This figure is best viewed in color. (RQ2)}  
	    %\reduceSpaceAroundFigure
		\label{fig:rt_user_volume_top5srcs_pop}
	\end{figure}

	Figure~\ref{fig:rt_user_volume_top5srcs_pop} illustrates the  distributions of the number of users who retweet each source tweet for the five most frequently retweeted sources of each type. We see a clear difference in the behavior of trusted sources and disinformation sources. Disinformation sources show a significant shift in the bulk of source tweets compared to the other source types. The same shift is seen in the most popular propaganda source. These sources have a greater proportion of their tweets retweeted by a larger number of users than the other types of sources, including trusted news. As expected, the number of users who retweet is closely correlated with the number of source tweets (Pearson = 0.76, $p<0.01$).

	\rqthreesection Next, we study how quickly users share direct retweets compared to tweets that @mention source accounts. We find that, as expected, {\it the majority of retweets occur within 24 hours of the original tweet being posted}, regardless of whether the share was a direct retweet or an @mention of a source. Although previous work found longer delays when deceptive content like rumor is propagated compared to verified news~\cite{jin2013epidemiological}, these trends did not appear for all types of deceptive sources when we considered all content posted, \ie deceptive sources may post both deceptive and non-deceptive content. 
	
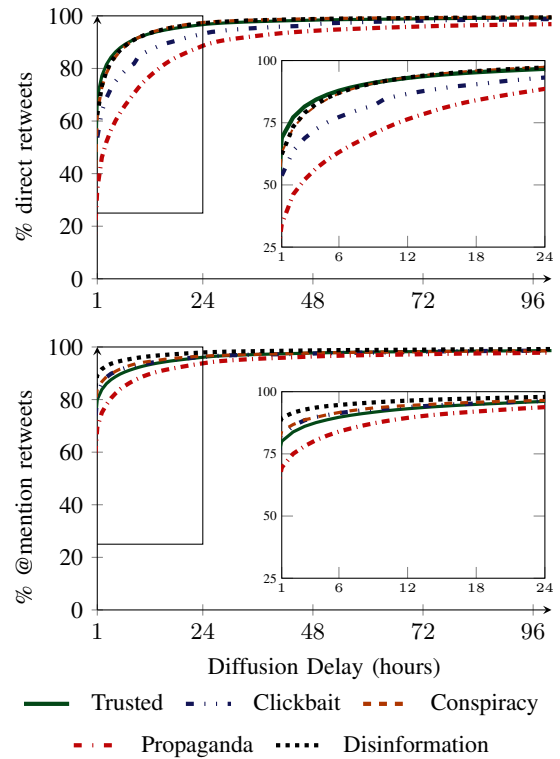
\begin{figure}[!bt]
		\centering     
		%\vspace{-.1in}
		\include{figs/journal/RTdiffusion_delay_cdfs}  
		\vspace{-.2in}%325in}
		\include{figs/journal/MTdiffusion_delay_cdfs} 
		\vspace{-.3in}
		\begin{tikzpicture} 
\begin{axis}[%
hide axis, height =.75in,
xmin=0,xmax=50,ymin=0,ymax=0.4,
legend style={at={(0.5,1)},%0.5)},
	anchor=north,legend columns=-1,column sep=.2cm,draw=none},   
]
\addlegendimage{no marks, V,ultra thick}
\addlegendentry{Trusted};%Verified};
\addlegendimage{no marks,CB,loosely dashdotdotted, ultra thick}
\addlegendentry{Clickbait};%Clickbait};
\addlegendimage{no marks,CS,dashed, ultra thick}
\addlegendentry{Conspiracy};%Conspiracy}; 
\end{axis}
\end{tikzpicture}

\begin{tikzpicture} 
\begin{axis}[%
hide axis, height =.75in,
xmin=0,xmax=50,ymin=0,ymax=0.4,
legend style={at={(0.5,1)},%0.5)},
	anchor=north,legend columns=-1,column sep=.2cm,draw=none},   
] 
\addlegendimage{no marks,P,loosely dashdotted, ultra thick}
\addlegendentry{Propaganda};%Propaganda};
\addlegendimage{no marks,D,dotted, ultra thick}
\addlegendentry{Disinformation};%Disinformation}; 
\end{axis}
\end{tikzpicture}
		\reduceSpaceBeforeCaption   
		\caption{
		CDF plots of diffusion delays (in hours) by news type for direct retweets (top) and retweets with @mentions of the source account (bottom). The inset of each plot shows a closer view of the initial diffusion, highlighted with a box in the larger plot. (RQ3)}
		\label{fig:diffusion_delay_cdf_pop}
		%\reduceSpaceAroundFigure
	\end{figure}

	Delays of retweets from suspicious sources are, on average, longer than for trusted sources; however, their 95\% confidence intervals overlap. When we examine cumulative distribution functions (CDFs) of delays for each source type, shown in Figure~\ref{fig:diffusion_delay_cdf_pop}, we observe some statistically significant differences (MWU $p<0.01$) in how long users wait before they retweet from a specific type of deceptive source. Users retweet from trusted, conspiracy, and disinformation sources after similarly short delays (soon after content is posted). However, users who retweet from clickbait sources wait significantly longer ($p<0.01$) after the sources post content. Those who retweet propaganda sources wait the longest after original postings ($p<0.01$).

	There are noticeable differences between @mention tweets and direct retweets. A much larger percentage of @mention tweets are shared within the first hour after the original posting occurs than the content retweeted directly from a source for all source types. This would also include content originally posted that was then retweeted by at least one other user before being retweeted again with the source account mentioned within the retweet, \eg through the use of \url{RT @source}.  
	Diffusion delays for tweets that @mention disinformation sources in the body of the tweet appear to skew toward shorter delays in the bottom plot of Figure~\ref{fig:diffusion_delay_cdf_pop} than those mentioning trusted sources. However, MWU tests found the distributions did not differ significantly.

	\rqfoursection  
	Finally, we look at \textit{who} shared content from sources at an aggregate level. For this, we compared the overlap of users across all types of news sources. Table~\ref{tab:overlap_pop} shows these intersections as a percentage of the set of users who shared content from the column's type of sources. This measurement approximates the likelihood that users who interacted with sources of the column's type also interacted with sources of the row's type. 
	
	We see the highest intersections with trusted sources for all types of deceptive sources. That is, 68\%, 60\%, 41\%, and 27\% of users who retweeted information from clickbait, conspiracy, propaganda, and disinformation sources, respectively, also shared information from trusted sources. Intuitively, this makes sense because the mainstream trusted sources are likely to post a more general or broad range of content than the other types of sources which may be more targeted toward a niche set of users or viewpoints.

\begin{table}[!t]
	\caption{
		Overlaps of user accounts who retweeted multiple news source types. $x$ in row $i$ and column $j$ means that $x$\% of users who retweeted $j$-sources also retweeted $i$-sources. (RQ4) 
	}
	%\reduceSpaceAroundFigure
	\small
	\centering
	\begin{tabular}{@{\hskip0pt}r@{\hskip6pt}rrrrr} 
		{}& T &  CB & CS & P & D \\
		%\hline
		Trusted (T)~ & \cellcolor{V!100} {\color{white!95!black}100} &\cellcolor{CB!68} 68& \cellcolor{CS!60}60& \cellcolor{P!41}41&\cellcolor{D!27}27 \\
		
		Clickbait (CB)~ & \cellcolor{V!1} 1 & \cellcolor{CB!100}{\color{white!95!black}100} & \cellcolor{CS!8}8 & \cellcolor{P!3}3 & \cellcolor{D!1}1 \\
		Conspiracy (CS)~ & \cellcolor{V!2} 2 & \cellcolor{CB!15}15 &\cellcolor{CS!100}{\color{white!95!black}100} & \cellcolor{P!7}7 & \cellcolor{D!3}3 \\
		Propaganda (P)~ & \cellcolor{V!7} 7 & \cellcolor{CB!39}39 & \cellcolor{CS!43}43 &\cellcolor{P!100}{\color{white!95!black}100} & \cellcolor{D!9}9\\
		Disinformation (D)~ & \cellcolor{V!6} 6 &\cellcolor{CB!17}17 &\cellcolor{CS!23}23 & \cellcolor{P!11}11 & \cellcolor{D!80}{\color{white!95!black} 100} 
	\end{tabular}
	%\reduceSpaceAroundFigure
	\label{tab:overlap_pop}
\end{table}

	We see an interesting overlap between propaganda, clickbait, and conspiracy sources. There is a high proportion of users who shared from clickbait-sources (39\%) or conspiracy-sources (43\%) who also propagated information from propaganda sources. However, relatively few users who shared from propaganda sources also shared from clickbait or conspiracy sources --- only 3\% and 7\%, respectively. Users who retweeted clickbait and conspiracy sources are fairly likely to have retweeted propaganda sources, but not the other way around. In fact, users who retweeted clickbait and conspiracy sources are the most likely to have also retweeted other source types.

	We then studied social graphs of users who directly interacted with each type of news sources. We compare network statistics of each set of graphs and report {\it key novel findings} below:
	\begin{itemize}[]%noitemsep]%,nolistsep]
	\item The density (\ie edge to node ratio) of propaganda networks are significantly different from trusted networks ($p<0.01$).
	\vspace{0.05in}
	\item The density, average in-degree, and average out-degree of trusted networks differ from conspiracy theory networks ($p<0.05$) and propaganda networks ($p<0.01$).
	\vspace{0.05in}
	\item Average shortest path lengths of disinformation networks differ significantly from trusted  ($p<0.01$), conspiracy theory ($p<0.05$), and propaganda networks ($p<0.05$).
	\vspace{0.05in}
	\end{itemize}

	\begin{table*}[t!]
	    \reduceSpaceAroundFigure 
		\caption{ 
		Gini coefficients and Palma ratios for each demographic attribute, averaged for each source type using source-user combinations as diffusion units. * denotes that a large majority ($>75\%$) of users were classified as the given demographic. Cells are highlighted based on the Gini coefficients (top) or the ratio of the Palma ratio to the perfect equality ratio of 0.25 (bottom). This figure is best viewed in color. (RQ1)} 
		\centering 
		\small
	    \reduceSpaceAroundFigure
		\setlength{\tabcolsep}{0.01em}%25em} 
		\begin{tabular}{lrr@{\hskip8pt}rr@{\hskip8pt}rr@{\hskip8pt}rr@{\hskip8pt}rr}
		
			\multicolumn{11}{l}{\sc Gini Coefficients} \\
			%Gini					
			
			{} & \multicolumn{2}{c}{Gender} & \multicolumn{2}{c}{Age} & \multicolumn{2}{c}{Income} & \multicolumn{2}{c}{Education} & \multicolumn{2}{c}{Role} \\  																
			\textit{(Source-Type)}&	\multicolumn{1}{r}{M*} 	&	\multicolumn{1}{c}{F} 	&	\multicolumn{1}{c}{Y} 	&	\multicolumn{1}{c}{O*} 	&	\multicolumn{1}{c}{B} 	&	\multicolumn{1}{c}{A*} 	&	\multicolumn{1}{c}{H} 	&	\multicolumn{1}{c}{C*}	&	\multicolumn{1}{c}{R}	&	\multicolumn{1}{c}{L} 	\\
			
			Trusted	& \cellcolor{yellow!30}{0.61} & \cellcolor{yellow!26} 0.52 & \cellcolor{yellow!22} 0.44 & \cellcolor{yellow!30} 0.61 & \cellcolor{yellow!31} 0.62  & \cellcolor{yellow!30} 0.61 & \cellcolor{yellow!29} 0.59 & \cellcolor{yellow!30} 0.61 & \cellcolor{yellow!31} 0.63  & \cellcolor{yellow!29} 0.58 \\
			Clickbait & \cellcolor{yellow!18} 0.37 	&	 \cellcolor{yellow!16} 0.32 	&	  \cellcolor{yellow!15} 0.31 	&	  \cellcolor{yellow!18} 0.37 	&	   \cellcolor{yellow!22} 0.45 	&	  \cellcolor{yellow!18} 0.36 	&	  \cellcolor{yellow!21} 0.43 	&	  \cellcolor{yellow!18} 0.36 	&	  \cellcolor{yellow!19} 0.38 	&	\cellcolor{yellow!18} 0.36 	\\
			
			Conspiracy & \cellcolor{yellow!34} 0.69 	& \cellcolor{yellow!17} 0.34 	& \cellcolor{yellow!23} 0.46 	&	  \cellcolor{yellow!34} 0.69 	&	  \cellcolor{yellow!36} 0.73 	&	  \cellcolor{yellow!33} 0.67 	&	  \cellcolor{yellow!36} 0.73 	&	  \cellcolor{yellow!33} 0.67 	&	  \cellcolor{yellow!35} 0.72 	&	 \cellcolor{yellow!32} 0.65	\\
			
			Propaganda & \cellcolor{yellow!28} 0.56 	&	\cellcolor{yellow!26} 0.52 	&	  \cellcolor{yellow!25} 0.51 	&	  \cellcolor{yellow!28} 0.56 	&	  \cellcolor{yellow!30} 0.61 	&	  \cellcolor{yellow!26} 0.54 	&	  \cellcolor{yellow!29} 0.59 	&	  \cellcolor{yellow!27} 0.55 	&	  \cellcolor{yellow!28} 0.57 	&	\cellcolor{yellow!27} 0.55	\\
			
			Disinformation~~~ & \cellcolor{yellow!43} 0.87 	&				\cellcolor{yellow!42} 0.84 	&	  \cellcolor{yellow!44} 0.88 	&	  \cellcolor{yellow!42} 0.84 	&	  \cellcolor{yellow!44} 0.88 	&	  \cellcolor{yellow!37} 0.75 	&	  \cellcolor{yellow!44} 0.88 	&	  \cellcolor{yellow!38} 0.78 	&	  \cellcolor{yellow!44} 0.88 	&	\cellcolor{yellow!42} 0.85 	\\ 
			
			\multicolumn{11}{l}{ } \\
			
			\multicolumn{11}{l}{\sc Palma ratios} \\
			{} & \multicolumn{2}{c}{Gender} & \multicolumn{2}{c}{Age} & \multicolumn{2}{c}{Income} & \multicolumn{2}{c}{Education} & \multicolumn{2}{c}{Role} \\  															
			\textit{(Source-Type)}&	\multicolumn{1}{r}{M*} 	&	\multicolumn{1}{c}{F} 	&	\multicolumn{1}{c}{Y} 	&	\multicolumn{1}{c}{O*} 	&	\multicolumn{1}{c}{B} 	&	\multicolumn{1}{c}{A*} 	&	\multicolumn{1}{c}{H} 	&	\multicolumn{1}{c}{C*}	&	\multicolumn{1}{c}{R}	&	\multicolumn{1}{c}{L} 	\\
			
			Trusted	&	 \cellcolor{yellow!18} 4.47	&	 \cellcolor{yellow!12} 3.08	&	  \cellcolor{yellow!8} 2.18	&	\cellcolor{yellow!20} 4.48 &  \cellcolor{yellow!20} 4.67	&  \cellcolor{yellow!17} 4.35 		&	  \cellcolor{yellow!16} 4.09 	&	 \cellcolor{yellow!18} 4.49 		&	  \cellcolor{yellow!19} 4.80		&	 \cellcolor{yellow!16} 4.00	\\
			Clickbait	&	  \cellcolor{yellow!7} 1.72	&	  \cellcolor{yellow!5} 1.40	&	  \cellcolor{yellow!5} 1.37	&	\cellcolor{yellow!7} 1.72				&	  \cellcolor{yellow!9} 2.38				&	  \cellcolor{yellow!7} 1.66				&	  \cellcolor{yellow!8} 2.19			&	 \cellcolor{yellow!7} 1.68		&	  \cellcolor{yellow!7} 1.79 	&	 \cellcolor{yellow!7} 1.65	\\
			Conspiracy	&	  \cellcolor{yellow!27} 6.58	&	  \cellcolor{yellow!5} 1.44	&	  \cellcolor{yellow!9} 2.28 	&	  \cellcolor{yellow!26} 6.58		&	 \cellcolor{yellow!35} 8.75		&	  \cellcolor{yellow!24} 6.00		&	  \cellcolor{yellow!36} 8.88		&	  \cellcolor{yellow!24} 6.02		&	  \cellcolor{yellow!31} 7.73	&	\cellcolor{yellow!21.48} 5.37	\\
			Propaganda	&	  \cellcolor{yellow!14} 3.58	&	  \cellcolor{yellow!12} 3.04	&	  \cellcolor{yellow!12} 2.91 	&	  \cellcolor{yellow!14} 3.58 	&	  \cellcolor{yellow!16} 4.41 	&	  \cellcolor{yellow!14} 3.39			&	  \cellcolor{yellow!16} 4.17			&	  \cellcolor{yellow!14} 3.45 	&	  \cellcolor{yellow!15} 3.74 				&	\cellcolor{yellow!14}3.42	\\
			Disinformation~~~	&	  \cellcolor{yellow!100} 25.06 	&	  \cellcolor{yellow!72} 18.03   	&	   \cellcolor{darkyellow!64!yellow} 41.34	&	  \cellcolor{yellow!72} 18.15 	&	  \cellcolor{darkyellow!40!yellow} 35.49	&	  \cellcolor{yellow!36} {\color{yellow!36}0}8.99  	&	  \cellcolor{darkyellow!42!yellow} 35.52	&	  \cellcolor{yellow!43} 10.66  	&	  \cellcolor{darkyellow!13!yellow} 28.26 	&	\cellcolor{yellow!84} 21.29 
		\end{tabular}  
		\label{tab:gini_palma_demog} 
	\end{table*}

	\section{Deception Propagation Results: \\ By Demographics}  
	
	In this section we report how evenly, how much, how quickly, and who shares content from news sources across and within each source type in context of the \textit{predicted user demographics} of our sample of 66,171 users who actively engage with deceptive news sources.

	\rqonesection  
	\noindent
	As we found at the population level, there is a relatively small group of users who share more than others. This is also reflected in the Gini coefficients and Palma ratios for all source type and demographic combinations in Table~\ref{tab:gini_palma_demog}. When we compare content shares from trusted (T) sources, we see the greatest differences between users in different age brackets. The biggest differences between sub-populations, however, occur between users who share from conspiracy and disinformation sources. Men who shared from conspiracy sources did so much more unevenly than women who shared from these sources, with an 84\% higher Palma ratio. A similar pattern occurs between users of different incomes; the Palma ratio is 2.95 times higher for users with higher incomes than users with incomes below \$35,000.

	As highlighted in Table~\ref{tab:gini_palma_demog}, the Palma ratios for disinformation sources are consistently higher than all other source types. {\it The greatest differences in equality at the extremes of the distribution of active users is seen in \textit{younger} users where the most active 10\% propagate 41.34 times as much as the least active 40\%.}  Disinformation sources are heavily retweeted by a slight proportion of the populations of younger users, users with lower incomes, and users with only a high school education who retweet a disinformation source at least once.

	\rqtwosection Next, we look at how many users within each demographic sub-population shared individual posts from sources of each type. In Table~\ref{tab:n_users_user_level}, we see that there are statistically significant differences (MWU $p<0.01$) in how many users retweeted each source post or post that @mentioned sources of each type. The demographic that had more users share each post is almost always the demographic which the majority of users were predicted to have (male, older, income above \$32,000, or college-educated). In one exception we find that users with income below \$35,000 (in 81\% of comparisons) or with only a high school education (in 75\% of comparisons) comprise the dominant share of users who shared individual posts from disinformation sources, despite there being fewer users predicted to have these attributes.

	\rqthreesection We compare the speed with which each demographic shares content posted by each type of source. CDFs of diffusion delays for each demographic (not illustrated) resulted in plots similar to those found in Figure~\ref{fig:diffusion_delay_cdf_pop}. Table~\ref{tab:diffusion_delay_user_level} illustrates which demographic takes a longer time to propagate content.

\begin{table}[!ht]
	\centering
	\small
	\caption{
	Demographic where more users shared each individual source post from sources of a given type with the common language effect size (as \% of comparisons) for each attribute. Statistical significance from MWU tests of $p<0.01$ for all results. (RQ2)}
	%% Users primarily labelled Male, Older, Above, Degree
	%\setlength{\tabcolsep}{0.3em} 
	%\reduceSpaceAroundFigure 
	\begin{tabular}{l|ccccc}%rrrrr} 
		%\multicolumn{6}{l}{\sc Direct retweets from news sources} \\
		Source-Type &                \multicolumn{1}{c}{Gender} &                   \multicolumn{1}{c}{Age} &                \multicolumn{1}{c}{Income} &             \multicolumn{1}{c}{Education} &                  \multicolumn{1}{c}{Role} \\ \hline
		Trusted    &   M 94 & O 98 & A 61 & C 68 &  W 34 \\
		Clickbait   &   M 94 & O 96 & A 70 & C 74 &  R 40 \\
		Conspiracy   &   M 99 & O 98 & A 56 & C 58 &  W 41 \\
		Propaganda    &   M 93 & O 90 & A 58 & C 63 &  R 40 \\
		Disinformation    &   M 96 & O 43 & B 81 & H 75 &  W 45 \\  
		 
	\end{tabular}
	
	%\reduceSpaceAroundFigure
	\label{tab:n_users_user_level}
\end{table}

	\begin{table}[!t]
		%\vspace{-.1in}
		%\centering
		\small
		\caption{
		Users by demographics who directly retweet from or @mention news source types after a longer delay and common language effect size (as \%). A dash (---) is shown if no significant differences were found. All results are statistically  significant (MWU tests of $p<0.01$). (RQ3)}
		\begin{tabular}{l}
		\textit{a. Direct retweets from news sources}
		\end{tabular}
		\begin{tabular}{l|ccccc}  
			 Source-Type &  Gender & Age & Income & Education & Role \\ 
			\hline
			Trusted         &   F 53    &   Y 56  &  A 52   &   C 51    &   R 49  \\
			Clickbait       &  ------   &  ------ &  A 54   &   C 52    &   L 52  \\
			Conspiracy      &   F 58    &   O 55  &  A 62   &   C 61    &   L 51  \\
			Propaganda      &   F 52    &   O 57  &  A 56   &   C 55    &   R 52  \\
			Disinformation~&   F 51    &   O 50  &  A 56   &   C 52    &   L 50  \\   
		\end{tabular}   
		\begin{tabular}{l}
		\\
		\textit{b. Retweets with @mentions of news sources}
		\end{tabular}
		\begin{tabular}{l|ccccc}  
			 Source-Type &  Gender & Age & Income & Education & Role \\ 
			\hline
			Trusted         &   F 54    &   Y 53  & ------  &  ------   &   R 56  \\
			Clickbait       &   F 54    &   Y 58  &  B 59   &   H 59    &  ------ \\
			Conspiracy      &  ------   &  ------ & ------  &  ------   &   R 53  \\
			Propaganda      &   F 53    &  ------ &  A 53   &   C 53    &   R 52  \\
			Disinformation~&   F 54    &   O 56  &  A 56   &   C 56    &   R 50  \\   
		\end{tabular}  
		 
        \reduceSpaceAroundFigure
		\label{tab:diffusion_delay_user_level}
	\end{table}

	Except for comparisons between predicted gender or age brackets for users who retweet clickbait sources, we find significant differences in diffusion delays for direct retweets for all other news source types. Users predicted to have only high school education directly retweet news sources of all types faster than those with a college education. Men retweet from sources more quickly than women for all source types except clickbait. Users with predicted income below \$35,000, predicted to have only a high school education, or who are ``regular'' users share content from clickbait sources faster than their counterparts. Interestingly, we see that while older users will retweet trusted sources more quickly than younger users, there is a greater delay when they retweet the most suspicious sources --- conspiracy, propaganda, or disinformation, relative to the delays of younger users.
	
	On the other hand, we see fewer occurrences of significant differences between sub-populations when users retweet content that only @mention the source rather than directly retweet it. Users with different predicted genders and age brackets now share tweets that @mention clickbait sources after different delays. Users predicted to be men, older, college-educated, or to have incomes below \$35,000 will share tweets that @mention clickbait sources faster than their counterparts, but only slightly (effect size of 54-59\%).

	\rqfoursection Finally, we study \textit{who} is sharing in terms of the predicted user demographics. Table~\ref{tab:mw_table_more_likely_to_propagate} presents the demographic that is more likely to directly retweet each source type at least once. We see that older users are less likely than younger users to share content from a disinformation source, but more likely for any other type of news. Similar patterns occur in users predicted to have higher versus lower income or education levels (who may also be older). When we consider the predicted gender, we see a similar trend where women are more likely to share content at least once from disinformation sources and less likely than men for all other types of suspicious sources. However, women are no less or more likely than men to share content from trusted sources.  
	We did not find any significant differences in demographics for @mentions. 

	\begin{table}[!ht]
	
		\caption{
		Demographics more likely to share a direct RT from a source of a given type at least once. For each attribute, the demographic with the higher likelihood is listed if statistically significant (MWU $p<0.01$) with the common language effect size (as \%). A dash (---) stands for cells if no significant difference is found. (RQ4)}
		
		\centering
		\small   
		\begin{tabular}{l|ccccc}% rrrrr}  
			 
			Source-Type &                \multicolumn{1}{c}{Gender} &                   \multicolumn{1}{c}{Age} &                \multicolumn{1}{c}{Income} &             \multicolumn{1}{c}{Education} &                  \multicolumn{1}{c}{Role} \\ \hline
			Trusted    &  F 26 &  O 26 &   B 24 &   H 24 &  L 23 \\
			Clickbait  &  M 05 &  O 05 &   A 05 &   C 05 & ------  \\
			Conspiracy   &  M 14 &  O 15 &   A 14 &   C 14 &  L 14  \\
			Propaganda    &  M 36 &  O 37 &   A 33 &   C 32  &  L 26   \\
			Disinformation    &  ------ &   Y 32 &  B 28   &   H 28  &  L 15  
			 
		\end{tabular} 
		\vspace{-1.4\baselineskip}
		\label{tab:mw_table_more_likely_to_propagate}
	\end{table}
	
	 \section{Summary}
	 	 Our extensive large-scale population-level and demographics analysis of the propagation behavior of users who directly interact with different types of news sources identified several key differences. Some characteristics, like diffusion inequality and the number of users who retweet per post, show large differences between trusted and disinformation news sources. Other results highlight key differences between propaganda, clickbait, and conspiracy news sources. Together, these novel results may be used to differentiate news sources of varying degrees of credibility without the need for expensive content-level annotation.
	  
	Recall that this paper explores four research questions at the population-level and across various demographics. A summary of our novel findings is presented below.

	\smallskip
	\noindent \textbf{\rqone}
	\begin{itemize} 
	\item Population: Direct retweets of disinformation sources are most highly retweeted from a small group of users that actively engage with those sources regularly. Propaganda is the next most unevenly shared news, followed by trusted news, conspiracy, and clickbait. \smallskip
	\item By demographics: We find striking differences in sharing behavior across different user demographics. The largest imbalance is in the most active 10\% of young users, which retweet disinformation sources 41.34 as much as the least active 40\%.
	\end{itemize}
	
	\smallskip 
	\noindent \textbf{\rqtwo}
	\begin{itemize} 
	\item Population: We did not find statistically significant differences between the number of users that retweet suspicious news across source types. The exception to this finding was disinformation sources, which have a higher number of users who retweet each post compared to trusted sources. \smallskip
	\item By demographics: Users, on average, with an annual income below \$35,000 (in 81\% of comparisons) or high school-educated users (in  75\% of comparisons) share content from disinformation sources more often than their counterparts.
	\end{itemize}
	
	\smallskip 
	\noindent \textbf{\rqthree}
	\begin{itemize} 
	\item Population: Trusted, conspiracy, and disinformation sources  have similarly short delays between the time a source posts content and the time that users share it. Delays are significantly longer for clickbait and propaganda sources ($p<0.01$).
	\smallskip
	\item By demographics: Older users retweet trusted sources more quickly than younger users. Younger users share the most suspicious sources (conspiracy, propaganda, and disinformation) more quickly than older users.
	\end{itemize}
	  
	\smallskip 
	\noindent \textbf{\rqfour}
	\begin{itemize} 
	\item Population: Users who share information from clickbait and conspiracy news sources are also likely to also share from propaganda sources, but not the other way around. \smallskip
	\item By demographics: Users who are within the majority predicted demographics are more likely to share at least once from all types of sources except disinformation, where the minority demographics is more likely to share. We found no significant results for which demographics are more likely to @mention sources.
	\end{itemize}

     \vspace{-.2\baselineskip}
	 \section{Limitations}
	 It is important to highlight some of the limitations in our study. First, the data samples used were not random samples nor representative of the overall Twitter population. It is important to note that we do not make claims about the behavior of \textit{all Twitter users}. We instead focus on the behavior of users who share information from deceptive news sources identified as conspiracy, propaganda, clickbait and disinformation. 
	 Second, the demographic labels were assigned by an imperfect classifier (even though state-of-the-art performance has been achieved). Some of the demographic classes had AUROC rates of 0.72. It is unclear if there is a classification bias in one direction or another. Nevertheless, we caution the reader against making strong claims for individual demographic classes.
	  
	 \section{Conclusions and Future Work}
	 To summarize, this is the first study that reports novel observable differences of information spread at the \textit{news account level} used to understand re-sharing behavior from sources of varying degrees of credibility in social media. More specifically, this work quantifies how people share misleading, manipulated, or potentially fabricated information from social media news sources; who these  propagators are; and how much and how evenly deceptive information is being shared. 
	 The properties we highlight in the previous section are differences in the way users directly interact with news sources that could be used to differentiate sources of varying credibility or trustworthiness without the need for tweet level annotation of deceptive versus trusted content or third-party source annotations.
	 
	 The results of our findings can be used to inform many practical applications including but not limited to: informing models and simulations of deceptive content spread across languages, geolocations, specific groups of users with different demographics and interests e.g., gatekeepers or persistent groups, and types of content e.g., deceptive posts during natural disasters or health messaging campaigns. These models can in turn be used to identify sources of varying credibility, or sources which require further investigation into credibility. For example, such a model that tags trusted versus deceptive, or potentially deceptive, sources could be used to guide not only the general public when they consume information from social news sources but journalists and fact-checkers who focus on verifying news sources and information being spread.
	 
	 Our analysis focuses on first-hop spreaders of deceptive news content in one social media platform and could naturally be extended to measure how information spreads from deceptive news sources beyond the first hop across many social environments. One could construct information cascades initiated by immediate-hop spreaders and measure how deceptive news propagate. For example, how deep deceptive news travel, how broad they go, how many unique users they reach, how many total users they affect, how long does it take for them to reach the audience of a certain size, how deceptive news evolves while being re-shared, or what the mechanisms of re-sharing are e.g., retweets, quotes, comments etc. Another interesting application could be to evaluate how the same deceptive news content, that is potentially seeded by adversaries, propagates across different social platforms e.g., Twitter, Reddit vs. Facebook. Moreover, understanding the intent behind misleading and fabricated news spread is another practical application of our analysis, that goes beyond rumor propagation work. In general, analyzing different types of online social behavior relevant to information spread e.g., information campaigns, coordinated effort, competing campaigns, recurrence, intimidation campaigns etc. is not only critical for national security but would also ensure healthier interactions and boost the level of trust in social media.

\section*{Acknowledgment}
	 The research described in this paper was supported by the Laboratory Directed Research and Development Program at Pacific Northwest National Laboratory, a multiprogram national laboratory operated by Battelle for the U.S. Department of Energy. Twitter data used for the analysis in this paper was acquired from a vendor and analyzed over the period of 01/2016 -- 01/2017.



\bibliographystyle{IEEEtran}
\bibliography{deception_propagation_bibfile}
%\begin{thebibliography}{1}


%\end{thebibliography}

\newpage

\begin{IEEEbiography}[{\includegraphics[width=1in,height=1.25in,clip,keepaspectratio]{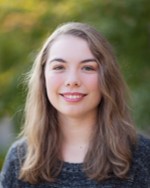}}]{Maria Glenski} received the M.S. degree from University of Notre Dame, Notre Dame, IN,
USA,
in 2018. She is currently pursuing the Ph.D. degree with the Department of Computer Science and Engineering,
University of Notre Dame, Notre Dame, IN,
USA.

She is also a member of the Interdisciplinary Center for Network Science and Applications (iCeNSA) at the University of Notre Dame. Her research in social media analysis and rating systems has been published in the ACM Conference on Hypertext and Social Media, ACM Transactions on Intelligent Systems and Technology, and the ACM Conference on Computer-Supported Cooperative Work and Social Computing.

\end{IEEEbiography}

\begin{IEEEbiography}[{\includegraphics[width=1in,height=1.25in,clip,keepaspectratio]{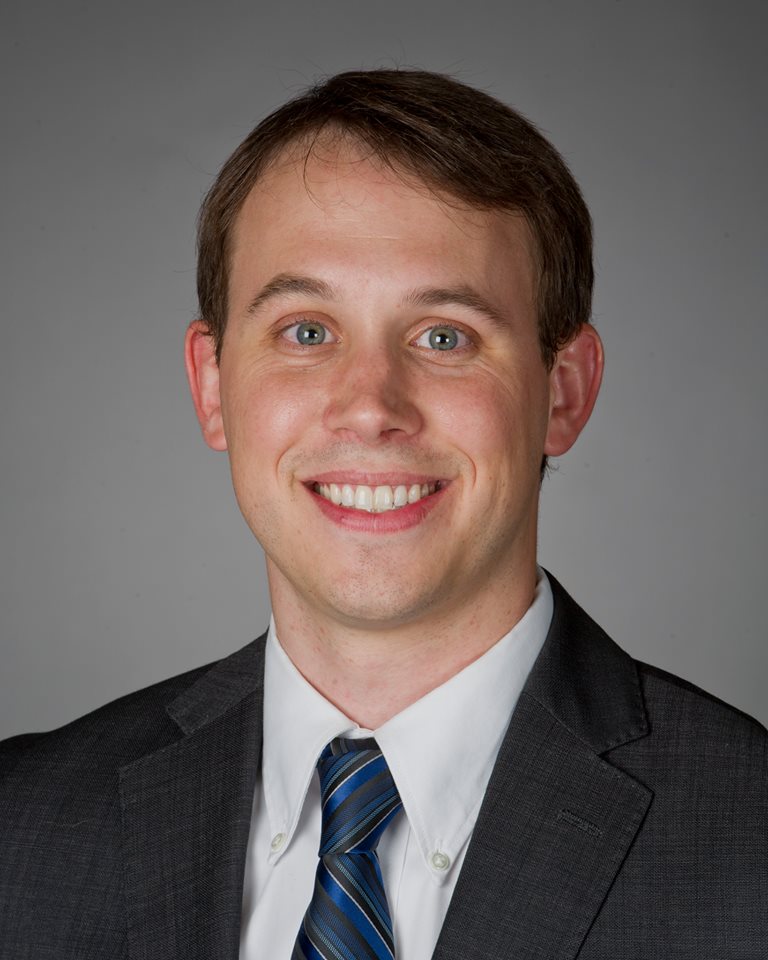}}]{Tim Weninger} received the Ph.D. degree from the University of Illinois Urbana–-Champaign, Champaign, IL, USA, in 2013. 

He is currently an Assistant Professor with the Department of Computer Science and Engineering, University of Notre Dame, Notre Dame, IN, USA, where he is also a member of the Interdisciplinary Center for Network Science and Applications. His current research interests include the intersection of machine learning, data mining, and network science in which he studies how humans create and consume networks of information. He has received research grants from NSF, AFOSR, and the John Templeton Foundation.
 
 Dr. Weninger has served on the program committee for numerous conferences and journals.

\end{IEEEbiography}

\begin{IEEEbiography}[{\includegraphics[width=1in,height=1.25in,clip,keepaspectratio]{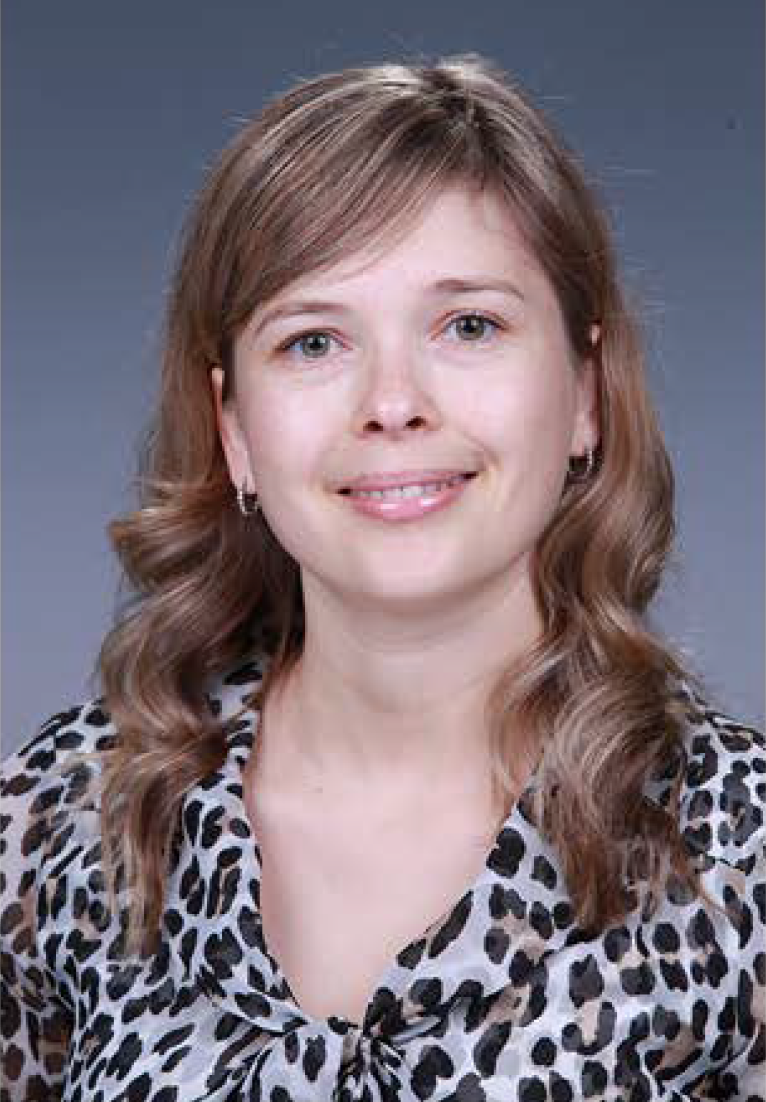}}]{Svitlana Volkova} received the Ph.D. degree from Johns Hopkins University, Baltimore, MD, USA, in 2015.

She is currently a senior scientist at the Data Sciences and Analytics group, National Security Directorate, Pacific Northwest National Laboratory. Her research focuses on advancing machine learning, deep learning and natural language processing techniques to build novel predictive and forecasting social media analytics. Her models advance understanding, analysis, and effective reasoning about extreme volumes of dynamic, multilingual, and diverse real-world social media data. She was awarded the Google Anita Borg Memorial Scholarship in 2010 and the Fulbright Scholarship in 2008. 

Dr. Volkova is a Vice Chair of the ACM Future of Computing Academy. 
\end{IEEEbiography}

\end{document}